\documentclass[doublecolumn,10pt]{IEEEtran}
\IEEEoverridecommandlockouts

\usepackage{amsfonts,amssymb,amsmath}
\usepackage{graphicx}
\usepackage{epsfig,array,cite}
\usepackage{epsf,exscale,times}
\usepackage{color,multicol}
\usepackage{booktabs,caption}
\usepackage[flushleft]{threeparttable}
\usepackage{algcompatible}
\usepackage[linesnumbered, ruled]{algorithm2e}
\usepackage[noend]{algpseudocode} 
\usepackage{footnote}
\usepackage{slashbox}
\usepackage{booktabs,caption}
\usepackage[flushleft]{threeparttable}
\usepackage{subcaption, comment, bm}
\usepackage[hang,flushmargin]{footmisc}
\usepackage{lipsum}

\let \bd = \textbf

\let \t  = \text

\def \p1#1{#1^{-1}}

\def \~#1{\tilde{#1}}

\let \nn = \nonumber

\algnewcommand\INPUT{\item[\textbf{Input:}]}%
\algnewcommand\OUTPUT{\item[\textbf{Output:}]}%
\begin{document}
%
\title{Adaptive Sparse Array Beamformer Design by Regularized Complementary Antenna Switching}

\author{
\IEEEauthorblockN{Xiangrong Wang\IEEEauthorrefmark{1}, Maria Greco\IEEEauthorrefmark{2}, Fulvio Gini\IEEEauthorrefmark{2}}\\
\IEEEauthorblockA{\IEEEauthorrefmark{1}School of Electronic and Information Engineering, Beihang University, Beijing, China; xrwang@buaa.edu.cn}
\IEEEauthorblockA{\IEEEauthorrefmark{2}Department of Information Engineering, University of Pisa, Pisa, Italy; m.greco@iet.unipi.it; f.gini@ing.unipi.it}
\thanks{The work by X Wang is supported by National Natural Science Foundation of China under Grant No. 62071021 and No. 61827901. The work of M. Greco and F. Gini has been partially supported by Italian Ministry of Education and Research (MIUR) in the framework of the CrossLab project (Departments of Excellence) of the University of Pisa, laboratory of Industrial Internet of Things (IIoT).}
}

\maketitle

\begin{abstract}

In this work, we propose a novel strategy of adaptive sparse array beamformer design, referred to as regularized complementary antenna switching (RCAS), to swiftly adapt both array configuration and excitation weights in accordance to the dynamic environment for enhancing interference suppression. In order to achieve an implementable design of array reconfiguration, the RCAS is conducted in the framework of regularized antenna switching, whereby the full array aperture is collectively divided into separate groups and only one antenna in each group is switched on to connect with the processing channel. A set of deterministic complementary sparse arrays with good quiescent beampatterns is first designed by RCAS and full array data is collected by switching among them while maintaining resilient interference suppression. Subsequently, adaptive sparse array tailored for the specific environment is calculated and reconfigured based on the information extracted from the full array data. The RCAS is devised as an exclusive cardinality-constrained optimization, which is reformulated by introducing an auxiliary variable combined with a piece-wise linear function to approximate the $l_0$-norm function. A regularization formulation is proposed to solve the problem iteratively and eliminate the requirement of feasible initial search point. A rigorous theoretical analysis is conducted, which proves that the proposed algorithm is essentially an equivalent transformation of the original cardinality-constrained optimization. Simulation results validate the effectiveness of the proposed RCAS strategy.

\end{abstract}

\IEEEpeerreviewmaketitle


\section{Introduction}
\label{sec:introduction}

\subsection{Background}

Antenna arrays have found extensive use for several decades in diverse applications, such as radar, sonar, telescope and communications to list a few \cite{Brennan1973,Compton1978,Gershman1995,Veen2004,WangX2021}. Antenna arrays sample signals in spatial domain and weight coefficients are then assigned to the received signals to achieve spatial filtering, which makes it possible to receive the desired signal from a particular direction while simultaneously blocking the interferences from other directions\cite{pillai2012,VanTrees2004,Krim1996}. Spatial filtering can be generally classified into two types: deterministic and adaptive. The former is also referred to as beampattern synthesis, which focuses on synthesizing beampatterns with prescribed mainlobe width and reduced sidelobe levels\cite{Nai2010,Massa2011,Fuchs2012,Poli2013,Wang2014a,wang2021active}. Adaptive beamforming extracts noise characteristics and intruding interference statistics from the received data and forms nulls towards the interferences automatically\cite{Talisa2016,Li2006,Lei2010,deAndrade2015}. It has been demonstrated in \cite{Wang2014, Wang2018} that adjusting excitation weights only is not adequate to combat mainlobe interferences. Array configurations are important spatial degrees of freedom (DoFs) for separating sources from interferences and its optimality varies with environmental scenario. Driven by the importance of array configurations, research into sparse array design techniques tailored for different applications continue unabated.

Sparse arrays comprise a set of judiciously activated antennas that maximizes the performance, while lowering system overheads and reducing computational complexity. Compared with uniformly spaced arrays, sparse arrays are capable of either increasing spatial resolution by enlarging array aperture given a fixed number of antennas or significantly reducing the cost while maximally preserving the performance\cite{Joshi2009,Wang2014,Fulton2016,Amin2016,Rajamaeki2018,Rocca2016}.
Sparse arrays designed in terms of beampattern synthesis are termed as deterministic sparse arrays, the antenna positions of which are fixed once calculated off-line. Although deterministic sparse arrays are blind to the situation, they possess resilient though average interference suppression performance regardless of surrounding environment, and thus are good choices when no environmental information is available. Sparse arrays designed in terms of maximizing output signal-to-interference-plus-noise-ratio (MaxSINR) are referred to as adaptive sparse arrays, where a subset of different sensors are adaptively switched on in accordance with changing environmental situations. Therefore, adaptive sparse arrays are superior to deterministic arrays in the metric of interference suppression thanks to their desirable situational awareness and exclusive focus on strong interferences\cite{Wang2018,Wang2018a,Hamza2019,Nosrati2020}.



\subsection{Relevant Work on Sparse Array Design}

Deterministic sparse array design has been extensively investigated in the literature, which aims at synthesizing a desired beampattern with the smallest number of antennas using sparsity-promoting algorithms, such as reweighted $l_1$-norm \cite{Nai2010}, Bayesian inference \cite{Massa2011}, soft-thresholding shrinkage \cite{Wang2014a}, and other compressive sensing methods \cite{Fuchs2012,Nongpiur2014}. The optimality of deterministic sparse arrays is beampattern-specific, for example, optimal array configurations vary with the steering direction. Our previous work in \cite{Wang2018b, Wang2020} inspected the common deterministic sparse array design for multiple switched beams in both cases of fully and partially-connected radio frequency (RF) switch network. Fully-connected switch network can facilitate the inter-connection of any input port to all output ports \cite{Rodriguez2017}. While partially-connected switch network constrains regularized antenna switching, which divides the full array into contiguous groups and precisely one antenna is selected in each group to compose the sparse array. The regularization of switched antennas is more practical in terms of circuit routing, connectivity and array calibration\cite{Nosrati2018} and furthermore restricts the maximum inter-element spacing, which, in turn, reduces the unwanted sidelobes. Other types of deterministic sparse arrays, such as minimum redundancy arrays (MRAs) and nested arrays \cite{Pal2010}, are designed to enable the estimation of more sources than physical sensors. However, they might exhibit significantly low array gain and are not suitable for interference mitigation\cite{Pal2010,Guo2018,Qin2019}. Nevertheless, no work has examined the regularized splitting of a large array into separate deterministic sparse arrays with complementary configurations, while collectively spanning the full aperture, and having at the same time well-controlled quiescent patterns. The solution to this problem constitutes the first novel contribution of this work.

 
Although adaptive sparse arrays exhibit advantages in terms of interference suppression, their practical implementation has been impeded by the stringent requirement of timely update of environmental information sensed by all antennas. To address this impediment, our previous work adopted a two-step adaptive filtering strategy, where the sparse array optimal for environmental sensing is first switched on and a different sparse array for adaptive beamforming based on the sensed information is reconfigured in the second step. The work in \cite{Hamza2019} configured a fully-augmentable sparse array to estimate the received data correlation matrix corresponding to the full array aperture, and then reconfigured another sparse array along with weights simultaneously for beamforming. The subsequent work in \cite{Hamza2020} utilized matrix completion to interpolate all missing spatial lags, whereas the accuracy of matrix completion heavily affects the sparse array design. The controversy of these strategies is that neither sparse arrays optimal for environmental sensing nor fully-augmentable sparse arrays are good configurations for beamforming. Thereby, the desired signal might get lost in the sensing stage and difficult to recover it in the following stages. It is favoured that the system could maintain acceptable interference suppression and normal functionality even in the sensing stage and make swift array reconfiguration upon situational changes. The solution to this problem is the second novel contribution of this work.

\begin{figure}
	\centering
	\includegraphics[scale=0.5]{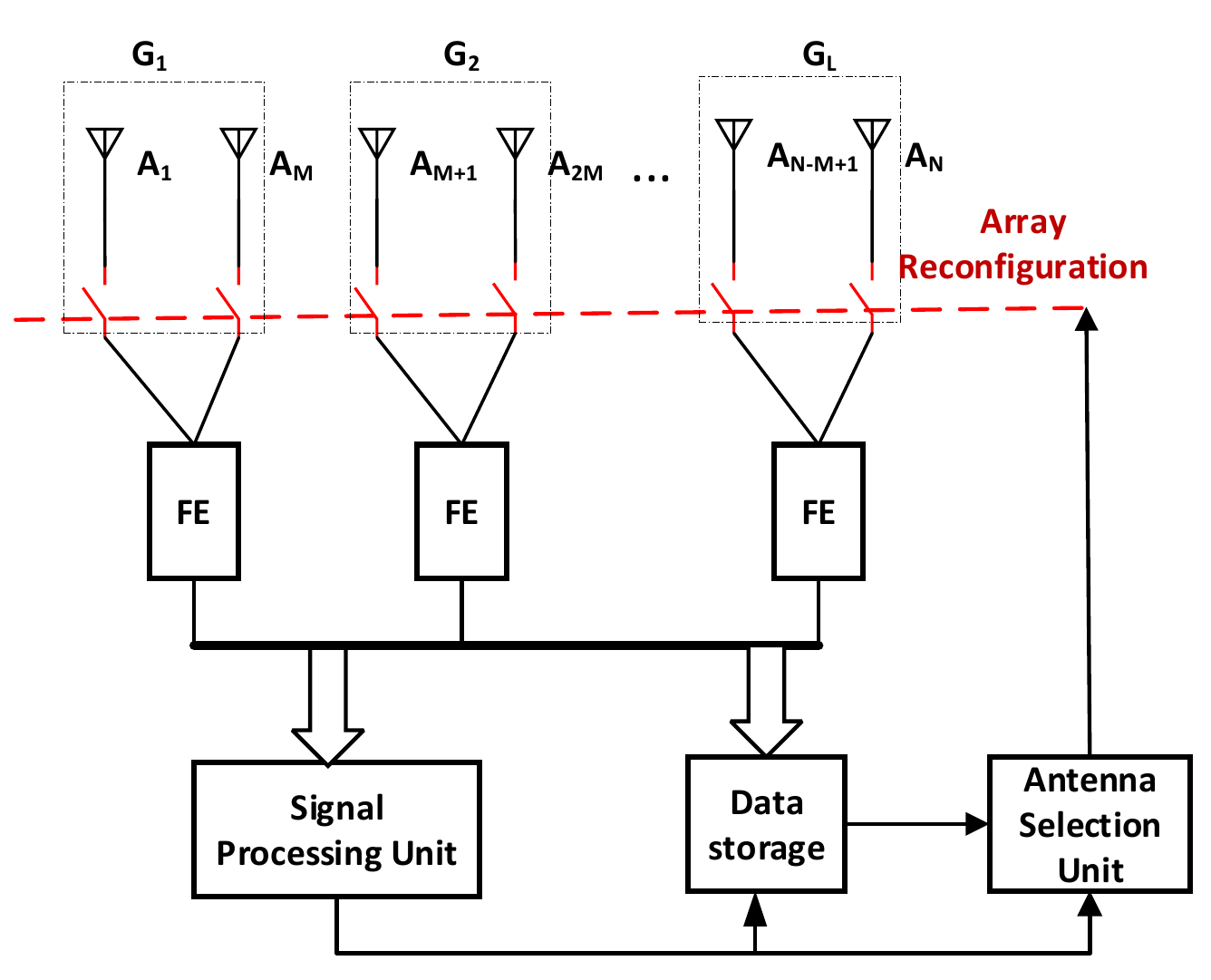}
	\caption{The hardware schematic of RCAS strategy: N antennas are divided into L groups and each group comprises M antennas. Only one antenna in each group is switched on.}
	\label{figch}
\end{figure}
\begin{figure}
	\centering
	\includegraphics[scale=0.5]{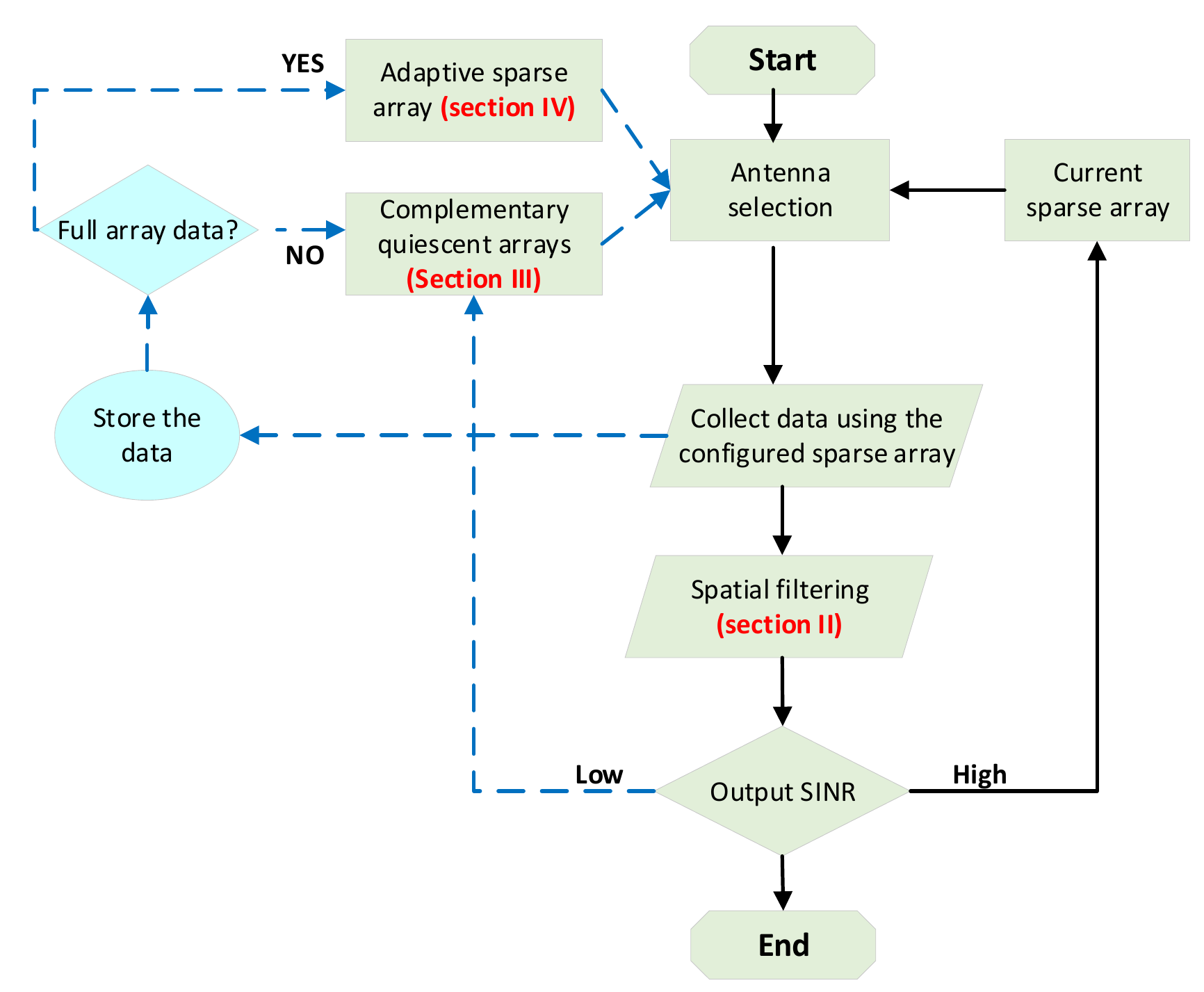}
	\caption{The workflow of the RCAS strategy: Complementary sparse arrays are sequentially switched on for environmental sensing, followed by adaptive sparse array design.}
	\label{figflow}
\end{figure}
%



\subsection{Proposed RCAS Strategy}

In this work, we investigate adaptive sparse array beamformer design and propose a novel regularized complementary antenna switching (RCAS) strategy. The difference between the proposed RCAS and our previous work is that (1) It does not require the assumption of known environmental information by alternate switching among complementary sparse arrays; (2) the filtering performance and normal functionality of the system are guaranteed even in the sensing stage; (3) regularized antenna switching is considered instead of unrestricted antenna selection to address practical circuitry issues and limitation on inter-element spacing. The hardware schematic of the RCAS is illustrated in Fig. \ref{figch}. A large array of $N$ uniformly spaced antennas is divided into $L$ groups and each group contains $M$ antennas. There is one front-end processing channel in each group, and the RF switch can connect any antenna in this group with the corresponding front-end within a sufficiently small time. The signal processing unit is responsible for spatial filtering and information extraction, and data storage will be started when the system is working in the sensing stage. Antenna selection unit calculates and determines the optimal array configuration in the specific environment and mandates the status change of the RF switches.

The proposed RCAS comprises two steps, deterministic complementary sparse array (DCSA) design, addressed in Section III, and regularized adaptive sparse array (RASA) design, addressed in Section IV. In the first step, the full array is divided into complementary sparse arrays by regularized antenna switching, all with good quiescent patterns and collectively spanning the full aperture. In the second step, the optimum sparse array for adaptive beamforming is then calculated and reconfigured based on the information sensed by all antennas. The workflow of the RCAS strategy is depicted in Fig. \ref{figflow}. In the case of ``cold start'', the set of complementary deterministic sparse arrays are first sequentially switched on for environmental sensing. The adaptive sparse array that is optimum in the specific environment can then be calculated and reconfigured, and spatial filtering is conducted to suppress the interferences. When spatial filtering experiences dramatic performance degradation (for example output SINR has dropped by 3dB), the currently employed adaptive sparse array loses its optimality for the specific environment. Complementary sparse arrays can be reconfigured back again for sensing the environmental dynamics, and the corresponding optimum adaptive sparse array needs to be re-designed. Note that spatial filtering, addressed in Section II, is continuously implemented on the configured sparse arrays even in the sensing stage such that the target of interest is always locked in focus.

In addition to the proposed RCAS strategy, we have made contributions to the methodologies embedded in the strategy, which are summarized as follows:
\begin{itemize}
\item
We introduce an auxiliary boolean variable combined with a piece-wise linear function to effectively approximate the cardinality constraint. 
\item 
We derive an affine upper bound to iteratively relax the non-convex constraints and a regularized formulation to eliminate the requirement of feasible start search point;
\item
We conduct theoretical analysis and prove that the proposed algorithm is an equivalent transformation of the original cardinality-constrained optimization.
\end{itemize}
%


In addition to the three main sections highlighted in Fig. 2, the rest of the paper is organized as follows: Simulation results, presented in section \ref{sec:simulations}, validate the effectiveness of proposed strategy. Finally, concluding remarks are provided in section \ref{sec:conclusion}.

\section{Review of Spatial Filtering Techniques}
\label{sec:review}

In this section, a full array is assumed and three beamforming techniques are reviewed as follows.

\subsection{Deterministic Beamforming}
\label{sec:deterministic}

Assume a linear array with $N$ antennas placed at the positions of $[x_1, \ldots, x_N]$ with an inter-element spacing of $d$. Deterministic beamforming essentially focuses on synthesizing a desired beampattern, which can be formulated as,
\begin{eqnarray}
\label{eq:expre_bs}
\min_{\bd w} \|\bd w^H \bd A_s - \bd f\|_2^2, \quad \t{s.t.} \; \bd w^H \bd a(\theta_0) = 1, 
\end{eqnarray}
where $\bd w \in \mathbb{C}^{N}$ is the $N$-dimensional weight coefficient vector with $\mathbb{C}$ denoting the set of complex number and $\bd a(\theta_0)$ is the steering vector towards the desired direction of $\theta_0$,
\begin{equation}
\bd a(\theta_0) = [e^{jk_0dx_1\sin\theta_0}, \ldots, e^{jk_0dx_N\sin\theta_0}]^T,
\end{equation}
where wavenumber is defined as $k_0= 2\pi /\lambda$ with $\lambda$ denoting wavelength and the steering direction $\theta_0$ is measured relative to the array broadside. Also, $\bd A_s \in \mathbb{C}^{N\times K}$ denotes the array manifold matrix of sidelobe region $\Omega_s = [\theta_1, \ldots, \theta_K]$, and $\bd f \in \mathbb{C}^{1\times K}$ represents the desired complex sidelobe beampattern, which can be expressed as the element-wise product between beampattern magnitude and phase. That is $\bd f = \bd f_d \odot \bd f_p$, where $\odot$ denotes element-wise product. The beampattern phase $\bd f_p$ can be fully employed as additional DoFs to enhance the beampattern synthesis and can be iteratively updated as,
\begin{equation}
\label{eq:expre_complexf}
\bd f_p^{(k+1)} = \bd f^{(k)} \oslash |\bd f^{(k)}|,
\end{equation}
where the superscript $^{(k)}$ denotes the $k$th iteration and the operator $\oslash$ denotes element-wise division. The complex beampattern in the $(k+1)$th iteration is updated as $\bd f^{(k+1)}=\bd f_d \odot \bd f_p^{(k+1)}$. It was proved in \cite{Wang2014a} that this updating rule will converge to a non-differential power pattern synthesis, i.e., $\min \| |\bd w^H \bd A_s| - \bd f_d\|_2$. Note that the beampattern $\bd f$ is the only row vector and others are all column vectors in this paper.

Define the following matrix and vector,
\begin{eqnarray}
\label{eq:expre_B1}
\bd B = \left[
\begin{array}{cc}
\bd f \bd f^H & -\bd f\bd A_s^H \\
-\bd A_s \bd f^H & \bd A_s \bd A_s^H
\end{array}
\right],
\end{eqnarray}
and $\tilde{\bd w} = [1,\bd w^T]^T$. Then Eq. (\ref{eq:expre_bs}) can be rewritten as,
\begin{eqnarray}
\label{eq:expre_bs1}
\min_{\tilde{\bd w}} \tilde{\bd w}^H \bd B \tilde{\bd w}, \quad \t{s.t.} \; \tilde{\bd w}^H \bd C = \bd g, 
\end{eqnarray}
where $\bd C=[\tilde{\bd a}(\theta_0), \bd e]$ with $\tilde{\bd a}(\theta_0)=[-1,\bd a^T(\theta_0)]^T$ is the extended steering vector, $\bd e \in \{0,1\}^{(N+1)\times 1}$ has a single one at the first entry and other $N$ entries being zero, and $\bd g=[0,1]^T$. The optimal weight vector can be obtained from Lagrangian multiplier method, that is,
\begin{equation}
\label{eq:expre_beamformquic}
\tilde{\bd w}_o = \bd B^{-1} \bd C (\bd C^H \bd B^{-1} \bd C)^{-1} \bd g.
\end{equation}
%

Essentially, a weak interference is reckoned to come from each angle in the sidelobe region by deterministic beamformer, which attenuates all of them equally and simultaneously by a factor of $\bd f_d$, while maintaining a unit directional gain towards the desired signal\cite{VanTrees2004, Krim1996}.

\subsection{Adaptive Beamfoming}

The well-known adaptive Capon beamformer aims to minimize the array output power while maintaining a unit gain towards the look direction\cite{VanTrees2004},
\begin{eqnarray}
\label{eq:expre_formulation}
\min_{\bd w} \bd w^H \bd R \bd w, \quad \t{s.t.} \; \bd w^H \bd a(\theta_0) = 1,
\end{eqnarray}
where $\bd R \in \mathbb{C}^{N \times N}$ is the covariance matrix of the received data by the employed antenna array and can be theoretically written as,
\begin{equation}
\label{eq:expre_Rtrue}
\bd R = \sigma_s^2 \bd a(\theta_0) \bd a^H(\theta_0) + \sum_{j=1}^J \sigma_j^2 \bd a(\theta_j) \bd a^H(\theta_j) + \sigma_n^2 \bd I,
\end{equation}
where $J$ is the number of interferences, and $\sigma_s^2$, $\sigma_j^2$ and $\sigma_n^2$ denote the power of source, the $j$th interference and white noise, respectively. The Capon beamformer weight vector is,
\begin{equation}
\label{eq:expre_weightcapon}
\bd w_o = \eta \bd R^{-1} \bd a(\theta_0),
\end{equation}
where $\eta = 1/\bd a^H(\theta_0) \bd R^{-1} \bd a(\theta_0)$. Capon beamformer emphasizes on the strong interferences in order to minimize the total output power, which, in turn, unavoidably lifts the sidelobe level in other angular regions as a result of energy suppression towards the interferences' direction.

\subsection{Combined Adaptive and Deterministic Beamforming}

Although adaptive beamforming is superior to deterministic beamforming in terms of MaxSINR, high sidelobes of adaptive beamforming constitute a potential nuisance, especially, when directional interferers are suddenly turned on. As such, it is crucial to combine the merits of two types of sparse arrays, those are MaxSINR and well-controlled sidelobes. Inspired by our previous work in \cite{Wang2017}, a combined beamformer with a regularized adaptive and deterministic objectives is formulated as follows,
\begin{eqnarray}
\label{eq:expre_regularized}
\min_{\bd w} \bd w^H \bd R \bd w + \beta \|\bd w^H \bd A_s - \bd f\|_2^2, \quad \t{s.t.} \; \bd w^H \bd a(\theta_0) = 1,
\end{eqnarray}
where the trade-off parameter $\beta$ is adjusted to control the relative emphasis between quiescent pattern and output power. Combining Eqs. (\ref{eq:expre_bs1}), (\ref{eq:expre_formulation}) and (\ref{eq:expre_regularized}) rewrites the formulation as, 
\begin{eqnarray}
\label{eq:expre_diar}
\min_{\tilde{\bd w}} \tilde{\bd w}^H \bd R_b \tilde{\bd w}, \quad \t{s.t.} \; \tilde{\bd w}^H \bd C = \bd g,
\end{eqnarray}
where $\bd R_b = \tilde{\bd R} + \beta \bd B$ with
\begin{equation}
\label{eq:expre_Rt}
\tilde{\bd R} = \left[
\begin{array}{cc}
0 & \bd 0_N^T \\
\bd 0_N & \bd R
\end{array}
\right].
\end{equation}
Here, $\bd 0_N$ is a N-dimensional vector of all zeros. The optimal weight vector is therefore,
\begin{equation}
\label{eq:expre_weight11}
\tilde{\bd w}_o = \bd R_b^{-1} \bd C (\bd C^H \bd R_b^{-1} \bd C)^{-1} \bd g.
\end{equation}
By comparing Eqs. (\ref{eq:expre_weight11}) with (\ref{eq:expre_beamformquic}), we can observe that combined beamformer is capable of maintaining well-controlled beampattern while suppressing the interferences by adjusting the trade-off parameter $\beta$. When $\beta$ is small, the combined beamformer prioritizes strong interferences over white noise for achieving MaxSINR. Otherwise, the minimization of white noise gain, that is sidelobe level, becomes a superior task.


\subsection{Sparse Array Beamformer Design}

As mentioned in the Sect. I.A, sparse array beamformer design techniques are generally divided into deterministic and adaptive techniques. Deterministic sparse array design aims to minimize the sidelobe level leveraging both array configuration and beamforming weights. No environmental information is required by the deterministic sparse array design, as it views all the sidelobe angular region equally important. Thus, deterministic sparse array is capable of suppressing the interferences from arbitrary directions, though the null depth may not be sufficient for MaxSINR. In a nutshell, deterministic sparse arrays possess resilient interference suppression performance regardless of surrounding environment. While adaptive sparse array is designed in terms of MaxSINR, which focuses on nulling strong interferences and thus mandates timely environmental information, such as the total number of interferences $J$ and their respective arrival angles $\theta_j, j=1,\ldots, J$. The environmental situation is usually unknown and needs to be estimated periodically, especially in dynamic environment. The lack of environmental information is regarded as an impediment to the optimum design of adaptive sparse arrays, which is solved by the proposed RCAS strategy in this work.

\section{Deterministic Complementary Sparse Array Design}
\label{sec:det}

In the first step of the proposed RCAS strategy, the full array is divided into separate complementary sparse arrays all with good quiescent patterns, as a result, switching among them can collect the data of full array while maintaining a moderate interference suppression performance. Note that the set of complementary sparse arrays are fixed once designed, regardless of environmental dynamics. We delineate the design of deterministic complementary sparse arrays in this section.

The $N$-antenna full array is split into $M$ sparse arrays by regularized antenna switching and each array consists of $L$ antennas. Note that these $M$ complementary sparse arrays do not have overlapping antennas and each array is capable of synthesizing a well-controlled quiescent pattern. The regularized array splitting can be formulated as,
\begin{eqnarray}
\label{eq:expre_splitting}
\min_{\bd w_1, \ldots, \bd w_M} && \sum_{m=1}^M \|\bd w_m^H \bd A_s-\bd f\|_2  \\
\t{s.t.} && \bd w_m^H \bd a(\theta_0) = 1, m=1,\ldots,M, \nn\\
         && \|\bd P_l \bd w_m\|_0 \leq 1, l=1,\ldots,L, m=1,\ldots,M \nn\\
         && \sum_{m=1}^M \|\bd q_i^T \bd w_m \|_0  \leq 1, i=1,\ldots, N\nn
\end{eqnarray}
where $\bd f \in \mathbb{C}^{1 \times K}$ is a row vector of the desired beampattern in the sidelobe angular region, and $\bd w_m \in \mathbb{C}^{N \times 1}, m=1,\ldots, M$ is a sparse weight vector with only $L$ non-zero coefficients corresponding to $L$ selected antennas of the $m$th sparse array. The group selection matrix $\bd P_l \in \{0,1\}^{M\times N}$ are all zeros except those entries of $\bd P_l(1,1+(l-1)M), \bd P_l(2,2+(l-1)M), \ldots, \bd P_l(M,lM)$. The vector $\bd q_i \in \{0,1\}^{N \times 1}$ are all zeros except for the $i$th entry being one. The last two constraints of Eq. (\ref{eq:expre_splitting}) are combined together to restrain that only one antenna in each group is selected for each sparse array and no overlapping antennas are selected for different sparse arrays. For the split $M$ sparse arrays, the same quiescent beampattern is desired that the main beams be steered towards $\theta_0$ with the prescribed sidelobes. For the sake of an easy exposition, we mostly omit the intended angle $\theta_0$ in the rest of the paper, unless we specifically refer to a particular angle.

Put the $M$ beamforming weight vectors into a matrix $\bd W=[\bd w_1, \ldots, \bd w_M]\in \mathbb{C}^{N \times M}$. Equivalently, problem in Eq. (\ref{eq:expre_splitting}) can be rewritten as,
\begin{eqnarray}
\label{eq:expre_splittingP0}
(P_0) \; \min_{\bd W} && \|\bd W^H \bd A_s-\bd F\|_{\t{F}}  \\
\t{s.t.} && \bd W^H \bd a = \bd 1_M, \nn\\
         && \|\bd P_l \bd W \bd c_m\|_0 \leq 1, l=1,\ldots,L, m=1,\ldots,M \nn\\
         && \|\bd q_i^T \bd W \|_0  \leq 1, i=1,\ldots, N\nn
\end{eqnarray}
where $\bd F=\bd 1_M\bd f$ with $\bd 1_M$ being a $M$-dimensional column vector of all ones, $\| \bullet \|_{\t{F}} $ is the Frobenius norm of a matrix, and $\bd c_m \in \{0,1\}^M$ is a column selection vector with all entries being zero except for the $m$th entry being one.

\begin{figure}
	\centering
	\includegraphics[scale=0.48]{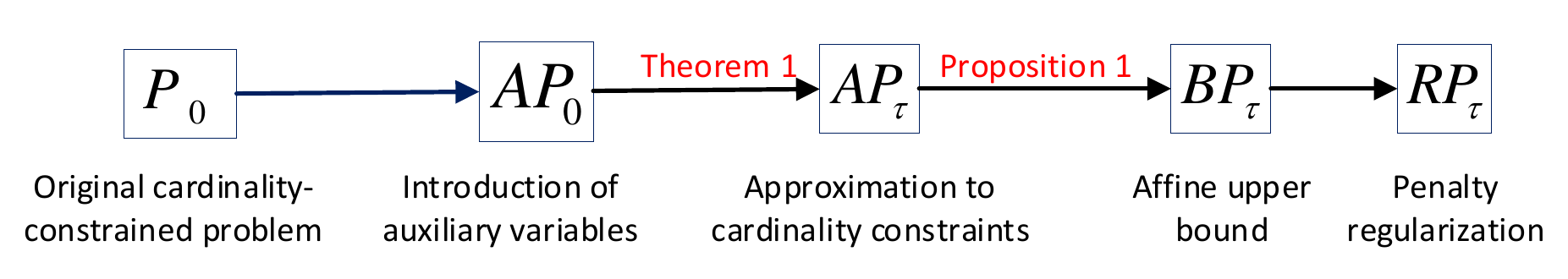}
	\caption{The derivation procedure of deterministic complementary sparse array design.}
	\label{figdev}
\end{figure}

The notorious cardinality constraints in the third and fourth lines of Eq. (\ref{eq:expre_splittingP0}) render the problem difficult to solve. In the following, we propose an algorithm to solve the cardinality-constrained optimization according to the derivation procedure illustrated in Fig. \ref{figdev}, where $P_0 \rightarrow AP_0$: an auxiliary variable is introduced to transform the complex-cardinality constraint to the real domain; $AP_0 \rightarrow AP_{\tau}$: a piece-wise linear function is utilized to approximate the $l_0$-norm; $AP_{\tau} \rightarrow BP_{\tau}$: an affine upper bound is derived to iteratively relax the non-convex constraints; $BP_{\tau} \rightarrow RP_{\tau}$: a regularized penalty is formulated to eliminate the requirement of feasible start search points.

\subsection{Introduction of Auxiliary Variables}
As the problem $(P_0)$ is formulated in the complex domain, we first introduce an auxiliary variable $\bd Z \in \mathbb{R}_+^{N \times M}$ to transform the complex cardinality constraint to the real domain. Here, $\mathbb{R}_+$ denotes the set of non-negative numbers. That is,
\begin{subequations}
\begin{eqnarray}
\label{eq:expre_splittinga}
(AP_{0}) \; \min_{\bd W, \bd Z} && \| \bd W^H \bd A_s - \bd F \|_{\t{F}}  \\
\t{s.t.} && \bd W^H \bd a = \bd 1_M, \nn\\
         \label{eq:expre_splittinga_line3}   
         && |\bd W| \leq \bd Z,\\ 
         \label{eq:expre_splittinga_line4}        
         && \|\bd P_l \bd Z \bd c_m \|_0 \leq 1, \\
         && l=1,\ldots,L, m=1,\ldots,M \nn \\
         \label{eq:expre_splittinga_line5}
         && \bd 1^T_M \bd P_l \bd Z \bd c_m =1, \\
         && l=1,\ldots,L, m=1,\ldots,M \nn \\    
         \label{eq:expre_splittinga_line6}     
         && \bd Z \bd 1_M  = \bd 1_{N},
\end{eqnarray}
\end{subequations}
where the inequality constraint in Eq. (\ref{eq:expre_splittinga_line3}) is element-wise, i.e., the absolute value of each element of $\bd W$ is bounded by the corresponding element of $\bd Z$. Constraints in Eqs. (\ref{eq:expre_splittinga_line4})-(\ref{eq:expre_splittinga_line6}) combined to restrain that different antennas are selected by the $M$ sparse arrays in each group. Furthermore, we can infer $\bd Z \in \{0,1\}^{N \times M}$ from the constraints in Eqs. (\ref{eq:expre_splittinga_line4}) and (\ref{eq:expre_splittinga_line5}), which implies that $\bd Z$ is an $N \times M$-dimensional matrix with entries being either 0 or 1. To put it differently, the binary property of the auxiliary variable $\bd Z$ is implicitly imposed in the formulation $(AP_{0})$. We can observe from problem $(AP_{0})$ that there is only one cardinality constraint left by introducing the auxiliary variable $\bd Z$.

Clearly, $\bd W$ solves $(P_0)$ if and only if $(\bd W,\bd Z)$ solves $(AP_{0})$ with $\bd Z = \t{sign}(|\bd W|)$, where $\t{sign}(\cdot)=0$ only when $\cdot = 0$, otherwise $\t{sign}(\cdot)=1$. The proof is as follows: (1) Suppose that $\bd W$ solves $(P_0)$, then set $\bd Z = \t{sign}(|\bd W|)$ and $(\bd W,\bd Z)$ solves $(AP_{0})$ as well. (2) Suppose that $(\bd W, \bd Z)$ solves $(AP_{0})$. When $Z_{ij}=1$, $W_{ij}$ must be non-zero for minimizing the objective function, whereas when $Z_{ij}=0$, we must have $W_{ij}=0$ for all $i=1,\ldots,N, j=1,\ldots,M$. Therefore, $\bd W$ solves $(P_0)$ as well. This proves the equivalence between $(P_0)$ and $(AP_0)$.

\subsection{Approximation to Cardinality Constraints}

As the $l_0$-norm is a notorious cardinality constraint and the problem involved is difficult to solve, we resort to a piece-wise linear function to approximate it. Assume that $\bd b \in \mathbb{R}_+^M$ is a $M$-dimensional non-negative real vector, and then the function is defined as,
\begin{equation}
\label{eq:expreb0}
\|\bd b\|_0 \approx \phi(\bd b,\bm{\tau}) = \sum_{i=1}^M (1/\tau_i) (b_i - (b_i - \tau_i)^+).
\end{equation}
where $\bm{\tau}=[\tau_1,\ldots,\tau_M]^T >0$ is a threshold vector, and $(b_i - \tau_i)^+ = \max\{b_i-\tau_i, 0\}$. Regarding to the properties of the approximation function  $\phi(\bd b, \bm{\tau})$, we have the following lemma.

\textbf{\textit{Lemma 1:}}

(1) For any $\bm{\tau} > 0$, $\phi(\bd b, \bm{\tau})$ is a piecewise linear under-estimator of $\|\bd b\|_0$, i.e. $\phi(\bd b, \bm{\tau}) \leq \|\bd b\|_0, \forall \bd b \in \mathbb{R}_+^M$, and $\phi(\bd b, \bm{\tau})$ is a non-increasing function of $\bm{\tau}$.

(2) For any fixed $\bd b \in \mathbb{R}_+^M$, it holds that
\begin{equation}
\lim_{\bm{\tau} \rightarrow 0^+} \phi(\bd b, \bm{\tau}) =  \|\bd b\|_0.
\end{equation}

(3) The sub-gradient of the function $\phi(\bd b,\bm{\tau})$ is $\bd g(\bd b,\bm{\tau}) = \frac{\partial \phi(\bd b,\bm{\tau})}{\partial \bd b}=[\frac{\partial \phi(\bd b,\bm{\tau})}{b_1}, \cdots, \frac{\partial \phi(\bd b,\bm{\tau})}{b_M}]^T$, where
\begin{equation}
\frac{\partial \phi(\bd b,\bm{\tau})}{b_i} = \begin{cases}
0, & \t{if} \; b_i > \tau_i \\
[0, 1/\tau_i],  & \t{if} \; b_i = \tau_i \\
1/\tau_i, & \t{if} \; b_i < \tau_i
\end{cases}
\end{equation}
and $[0,1/\tau_i]$ denotes any number between 0 and $1/\tau_i$. The proof of Lemma 1 is provided in Appendix A. 

Utilizing the approximation function of the $l_0$-norm in Lemma 1, problem $(AP_0)$ in Eq. (\ref{eq:expre_splittinga}) can be expressed as,
\begin{subequations}
\begin{eqnarray}
\label{eq:expre_splitting3}
(AP_{\tau}) \; \min_{\bd W, \bd Z} && \| \bd W^H \bd A_s - \bd F \|_{\t{F}}  \\
\t{s.t.} && \bd W^H \bd a = \bd 1_M, \nn\\
         \label{eq:expre_splitting3_line3}   
         && |\bd W| \leq \bd Z,\\ 
         \label{eq:expre_splitting3_line4}        
         && \phi(\bd P_l \bd Z \bd c_m,\bd P_l\bm{\Pi}\bd c_m) \leq 1, \\
         && l=1,\ldots,L, m=1,\ldots,M \nn\\
         \label{eq:expre_splitting3_line5}
         && \bd 1^T_M \bd P_l \bd Z \bd c_m =1, \\
         && l=1,\ldots,L, m=1,\ldots,M \nn\\    
         \label{eq:expre_splitting3_line6}     
         && \bd Z \bd 1_M  = \bd 1_{N}.
\end{eqnarray}
\end{subequations}
%
Here, $\bm{\Pi} \in \mathbb{R}_+^{N\times M}$ is the threshold matrix. The relationship between $(AP_0)$ and $(AP_{\tau})$ is given in Theorem 1.

\textbf{\textit{Theorem 1:}} When the threshold matrix $\bm{\Pi} <1$ (that is, $\Pi_{i,j} <1, \forall i=1,\ldots,N, j=1,\ldots,M$), the formulation $(AP_{\tau})$ is equivalent to the formulation $(AP_{0})$.

The proof of Theorem 1 is provided in Appendix B. 

\subsection{Affine Upper Bound}

As proved in Lemma 1, the approximation function $\phi(\bd b,\bm{\tau})$ is concave with respect to $\bd b$, thus the relaxed cardinality constraint in ($AP_{\tau}$) is still nonconvex. We then resort to the following affine function to iteratively upper-bound $\phi(\bd b,\bm{\tau})$,
\begin{equation}
\label{eq:expre_upbound}
\phi(\bd b,\bm{\tau}) \leq  \bar{\phi}(\bd b;\bd b_0,\bm{\tau}) = \phi(\bd b_0,\bm{\tau}) + \bd g^T(\bd b_0,\bm{\tau})(\bd b - \bd b_0),
\end{equation}
where $\bd g(\bd b_0,\bm{\tau})=[\frac{\partial \phi}{\partial b_1}, \ldots, \frac{\partial \phi}{\partial b_M}]^T\big |_{\bd b_0}$ denotes the sub-gradient of function $\phi(\bd b,\bm{\tau})$ evaluated at the point $\bd b_0$ and its definition is provided in the Lemma 1(3). Tracing back to problem ($AP_{\tau}$), the subgradient of the function $\phi(\bd P_l \bd Z \bd c_m,\bd P_l \bm{\Pi} \bd c_m)$ with respect to the variable $\bd Z$ is,
\begin{equation}
\label{eq:expre_subgradientZ}
\frac{\partial \phi(\bd P_l \bd Z \bd c_m,\bd P_l \bm{\Pi} \bd c_m)}{\partial \bd Z} = \bd P_l^T \tilde{\bd r}_{m,l} \bd c_m^T,
\end{equation}
where $\tilde{\bd r}_{m,l} = \sum_{i=1}^M g(b_i,\tau_i) \bd r_i$ with $b_i=\bd r_i^T \bd P_l \bd Z \bd c_m$ and $\tau_i=\bd r_i^T \bd P_l \bm{\Pi} \bd c_m$, and $\bd r_i \in \{0,1\}^{M \times 1}$ is a row selection vector with all entries being zero except for the $i$th entry. The proof of Eq. (\ref{eq:expre_subgradientZ}) is provided in Appendix \ref{subsec:proof_eq22}.

Substituting $\bd b = \bd P_l \bd Z \bd c_m$ and $\bd b_0 = \bd P_l \bd Z_0 \bd c_m$ into Eq. (\ref{eq:expre_upbound}) and utilizing the gradient in Eq. (\ref{eq:expre_subgradientZ}), we have that
\begin{eqnarray}
\label{eq:expre_phiapprox}
\phi(\bd P_l \bd Z \bd c_m,\bd P_l \bm{\Pi} \bd c_m) & \leq & \bar{\phi}(\bd P_l \bd Z \bd c_m; \bd P_l \bd Z_0 \bd c_m, \bd P_l \bm{\Pi} \bd c_m), \\
&=& \phi(\bd P_l \bd Z_0 \bd c_m,\bd P_l \bm{\Pi} \bd c_m)\nn\\
&+& \t{tr}\{(\bd c_m \tilde{\bd r}_{m,l}^T \bd P_l) (\bd Z - \bd Z_0)\}. \nn
\end{eqnarray}
Here, the variable changes from the vector $\bd b$ to the matrix $\bd Z$, thus the trace operator $\t{tr}(\cdot)$ is utilized instead of inner product.

Utilizing the upper bound in Eq. (\ref{eq:expre_phiapprox}) to Eq. (\ref{eq:expre_splitting3_line4}) yields,
\begin{eqnarray}
\label{eq:expre_splitting6}
(BP_{\tau}) \; \min_{\bd W, \bd Z} && \| \bd W^H \bd A_s - \bd F \|_{\t{F}}  \\
\t{s.t.} && \bd W^H \bd a = \bd 1_M, \nn\\
         && |\bd W| \leq \bd Z, \nn\\         
         && \phi(\bd P_l \bd Z_0 \bd c_m,\bd P_l \bm{\Pi} \bd c_m) \nn\\
         && +\t{tr}\{(\bd c_m \tilde{\bd r}_{m,l}^T \bd P_l) (\bd Z - \bd Z_0)\}\leq 1, \nn\\
         && l=1,\ldots,L, m=1,\ldots,M \nn\\
         && \bd 1^T_M \bd P_l \bd Z \bd c_m =1, l=1,\ldots,L, m=1,\ldots,M \nn\\         
         && \bd Z \bd 1_M  = \bd 1_N, \nn
\end{eqnarray}
%


The following proposition proves that the problem $(BP_{\tau})$ is equivalent to problem $(AP_{\tau})$.

\textbf{\textit{Proposition 1:}} When the initial search point $\bd Z^{(0)}$ for solving $(BP_{\tau})$ is feasible for $(AP_{\tau})$, the problem $(BP_{\tau})$ will converge to the problem $(AP_{\tau})$ iteratively.

The proof of Proposition 1 is shown in Appendix D. 

\subsection{Penalty Regularization}

The main drawback of the formulation $(BP_{\tau})$ is that the initial search point is required to be feasible to $(AP_{\tau})$. We next formulate an extension of $(BP_{\tau})$, which is capable of removing the requirement of an feasible start point. We add a penalty regularization by removing the upper bound of the piece-wise linear approximation function in the fourth line of Eq. (\ref{eq:expre_splitting6}) to the objective function, that is,
\begin{eqnarray}
\label{eq:expre_splitting7}
(RP_{\tau}) && \min_{\bd W, \bd Z} \| \bd W^H \bd A_s - \bd F \|_{\t{F}} + \rho \sum_{l=1}^L \sum_{m=1}^M \t{tr}\{(\bd c_m \tilde{\bd r}_{m,l}^T \bd P_l) \bd Z\} \nn \\
\t{s.t.} && \bd W^H \bd a = \bd 1_M, \\
         && |\bd W| \leq \bd Z, \nn\\         
         && \bd 1^T_M \bd P_l \bd Z \bd c_m =1, l=1,\ldots,L, m=1,\ldots,M \nn\\         
         && \bd Z \bd 1_M  = \bd 1_{N}, \nn
\end{eqnarray}
where $\rho$ is a predetermined parameter. Note that the regularized penalty in the objective of $(RP_{\tau})$ removes the constant terms $\phi(\bd P_l \bd Z_0 \bd c_m,\bm{\tau}) - \t{tr}\{(\bd c_m \tilde{\bd r}_{m,l}^T \bd P_l) \bd Z_0\}$ in the third constraint of $(BP_{\tau})$. Observe that for sufficiently large $\rho$, the second term in the objective function of $(RP_{\tau})$ becomes hard constraints and the two problems $(RP_{\tau})$ and $(BP_{\tau})$ are equivalent\cite{seq2010}. Assume that $\bd Z^{(k)}$ is the returned solution to $(RP_{\tau})$ in the $k$th iteration, and $\bd Z^{(k)}$ is infeasible to $(AP_{\tau})$. That is,
\begin{equation}
\bar{\phi}(\bd Z^{(k)}; \bd Z^{(k)},\bm{\Pi}) = \phi(\bd Z^{(k)},\bm{\Pi}) > 1.
\end{equation}
Since $\bar{\phi}(\bd Z; \bd Z^{(k)},\bm{\Pi})$ is decreasing iteratively for sufficiently large $\rho$, i.e., 
\begin{equation}
\label{eq:expre_sag1}
\bar{\phi}(\bd Z^{(k+1)}; \bd Z^{(k)},\bm{\Pi}) \leq \bar{\phi}(\bd Z^{(k)}; \bd Z^{(k)},\bm{\Pi}).
\end{equation}
The concavity of the function $\phi$ implies that 
\begin{equation}
\label{eq:expre_sag2}
\phi(\bd Z,\bm{\Pi}) \leq \bar{\phi}(\bd Z;\bd Z^{(k)},\bm{\Pi}).
\end{equation}
Combining Eqs.(\ref{eq:expre_sag1}) and (\ref{eq:expre_sag2}), we obtain that,
\begin{equation}
\label{eq:expre_sag3}
\phi(\bd Z^{(k+1)},\bm{\Pi}) \leq \phi(\bd Z^{(k)},\bm{\Pi}).
\end{equation}
We can argue from Eq. (\ref{eq:expre_sag3}) that the approximated cardinality function is decreasing iteratively. When $\phi(\bd Z^{(k)},\bm{\Pi})$ decreases to 1, $\bd Z^{(k)}$ becomes feasible to $(AP_{\tau})$, the problem $(RP_{\tau})$ will converge to the problem $(AP_{\tau})$ according to Proposition 1. 

\begin{algorithm}[!ht]
\caption{Deterministic Complementary Sparse Array (DCSA) Design} \label{alg:alg1}
  \SetKwInOut{Input}{Input}\SetKwInOut{Output}{Output}
  \Input{Parameter $\kappa=0.5$, $\zeta=0.001$, k=0, trade-off parameter $\rho$ and desired beampattern $\bd F=\bd F_d \odot \bd F_p$ with $\bd F_p=\bd 1$ and $\bd F_d$ specifies desired sidelobe level.}
  \Output{A set of complementary sparse arrays denoted by $\bd Z$.}
  \textbf{Initialize} the selection matrix $\bd Z^{(0)} \in [0,1]^{N \times M}$, \\
\Repeat{$\|\bd W^{(k+1)}-\bd W^{(k)}\|_{\t{F}}$ is sufficiently small}{
(1) Calculate the threshold matrix $\bm{\Pi}$,
\begin{equation}
\label{eq:expre_tauupdate}
\Pi_{i,j} = 
\begin{cases}
Z^{(k)}_{i,j} - \zeta, \; \text{if}\; Z^{(k)}_{i,j} \geq \kappa, \\
Z^{(k)}_{i,j} + \zeta, \; \text{if}\; Z^{(k)}_{i,j} < \kappa,
\end{cases}
\end{equation}
for $i=1,\ldots, N$ and $j=1,\ldots,M$.\\
(2) Calculate $\tilde{\bd r}_{m,l}, m=1,\ldots,M, l=1,\ldots,L$ according to the following formula,
\begin{equation}
\tilde{\bd r}_{m,l} = \sum_{i=1}^M g(z_i, \tau_i) \bd r_i, \nn
\end{equation}
where $z_i=\bd r_i^T \bd P_l \bd Z^{(k)} \bd c_m$, $\tau_i=\bd r_i^T \bd P_l \bm{\Pi} \bd c_m$ and,
\begin{equation}
g(z_i,\tau_i) = \begin{cases}
       0 & \t{if}\; z_i > \tau_i, \\
       1/\tau_i & \t{if}\; z_i \leq \tau_i. \nn
       \end{cases}
\end{equation}
(3) Solve problem $(RP_{\tau})$ based on $\bd Z^{(k)}$ to get $(\bd W^{(k+1)}, \bd Z^{(k+1)})$, set iteration number k = k+1, \\
(4) Update $\bd F_p = (\bd W^{(k+1)H}\bd A_s) \oslash |\bd W^{(k+1)H}\bd A_s|$ and  $\bd F = \bd F_d \odot \bd F_p$. \\
}
\end{algorithm}
%


We then propose deterministic complementary sparse array design in Algorithm \ref{alg:alg1}, which successively solves the problem $(BP_{\tau})$ and generates a sequence of $\{\bd W^{(k)}, \bd Z^{(k)}\}$. Proceeding from Eq. (\ref{eq:expre_complexf}), the updating formula of desired beampattern $\bd F$ is $\bd F_p = (\bd W^{(k+1)H}\bd A_s) \oslash |\bd W^{(k+1)H}\bd A_s|$ and $\bd F = \bd F_d \odot \bd F_p$. To better approximate the $l_0$-norm, the threshold $\bm{\Pi} \in \mathbb{R}_+^{N \times M}$ is updated iteratively according to $\bd Z$ in the previous iteration, per Eq. (\ref{eq:expre_tauupdate}) in line 3. The upper bound of the approximated cardinality constraints in $(BP_{\tau})$ is tight only in the neighbourhood of the initial point $\bd Z_0$, which renders the proposed algorithm a local heuristic\cite{Lipp2016}, and thus, the final solution depends on the choice of the initial point. We can therefore initialize the algorithm with different initial points $\bd Z^{(0)}$ randomly from the range $[0,1]$ and take the one with the lowest objective value over different runs as the final solution.




\section{Regularized Adaptive Sparse Array Design}
\label{sec:adaptive}

In the second step of the RCAS, the adaptive sparse array beamformer for the specific environment can then be calculated and reconfigured. The received data of the full array can be obtained by switching between the set of complementary sparse arrays designed in Sect. III. Specifically, when each sparse array is switched on, a length of $T$ samples are obtained and stored. After $MT$ sampling intervals, the data of the full array is obtained and stacked into a matrix $\bd Y \in \mathbb{C}^{N\times T}$. The covariance of the full array is then $\bd R_f=(1/T)\bd Y \bd Y^H$. Based on the covariance, the optimum sparse array for combined beamforming as revisited in Sect. II-C can be obtained by,
\begin{eqnarray}
\min_{\bd w} && \bd w^H \bd R_f \bd w + \beta \|\bd w^H \bd A_s - \bd f\|_2^2, \\
\t{s.t.} && \bd w^H \bd a(\theta_0) = 1, \; \|\bd P_l \bd w\|_0 \leq 1, l \leq 1,\ldots,L, \nn
\end{eqnarray}
where the second constraint is imposed to guarantee only one antenna is switched on in each group. Utilizing the same derivation procedure shown in Fig. 3, the problem is first transformed into the following formulation by introducing an auxiliary variable $\bd z$,
\begin{eqnarray}
\min_{\bd w,\bd z} && \bd w^H \bd R_f \bd w + \beta \|\bd w^H \bd A_s - \bd f\|_2^2, \\
\t{s.t.} && \bd w^H \bd a(\theta_0) = 1, \; |\bd w| \leq \bd z, \nn\\
         && \bd 1_M^T \bd P_l \bd z = 1, l = 1,\ldots,L, \nn\\
         && \|\bd P_l \bd z\|_0 \leq 1, l = 1,\ldots,L, \nn
\end{eqnarray}
where $\bd z \in \{0,1\}^N$ is the selection vector and bounds the absolute value of $\bd w$. Utilizing the piece-wise linear function to approximate the $l_0$-norm and upper-bounding the concave approximation function iteratively yield,
\begin{eqnarray}
\min_{\bd w,\bd z} && \bd w^H \bd R_f \bd w + \beta \|\bd w^H \bd A_s - \bd f\|_2^2, \\
\t{s.t.} && \bd w^H \bd a(\theta_0) = 1, \; |\bd w| \leq \bd z, \nn\\
         && \bd 1_M^T \bd P_l \bd z = 1, l = 1,\ldots,L, \nn\\
         && \phi(\bd P_l \bd z^{(k)}, \bd P_l \bm{\tau}) + \bd g^T(\bd P_l \bd z^{(k)}, \bd P_l \bm{\tau})\bd P_l(\bd z - \bd z^{(k)}) \leq 1, \nn\\
         && l = 1,\ldots,L \nn
\end{eqnarray}
where $\bd g(\bd P_l\bd z^{(k)}, \bd P_l\bm{\tau})$ is the gradient vector as defined in Lemma 1(3) and $\bm{\tau} \in \mathbb{R}_+^N$ is an $N$-dimensional threshold vector. Similar to $(RP_{\tau})$, we remove the third set of constraints as regularized penalties to the objective as follows,
\begin{eqnarray}
\label{eq:expre_adaptives}
\min_{\bd w,\bd z} && \bd w^H \bd R_f \bd w + \beta \|\bd w^H \bd A_s - \bd f\|_2^2 + \rho \bd g^T(\bd z^{(k)}, \bm{\tau}) \bd z \nn\\
\t{s.t.} && \bd w^H \bd a(\theta_0) = 1, \; |\bd w| \leq \bd z, \nn\\
         && \bd 1_M^T \bd P_l \bd z = 1, l = 1,\ldots,L,
\end{eqnarray}
%
%
where $\sum_{l=1}^L \bd g^T(\bd P_l \bd z^{(k)}, \bd P_l\bm{\tau})\bd P_l \bd z = \bd g^T(\bd z^{(k)}, \bm{\tau}) \bd z$ utilizing the definition of the matrix $\bd P_l, l=1,\ldots, L$. As complementary sparse arrays are deterministic design, it is acceptable to try different initial search points to secure the final optimum. For the design of adaptive sparse array, a good initial search point should be selected because of the requirement of real-time reconfiguration. The selection variable $\bd z$ can be initialized by a reweighted $l_1$-norm minimization\cite{Candes2008}. The optimization in the $k$th iteration is formulated as,
\begin{eqnarray}
\label{eq:expre_reweight}
\min_{\bd w,\bar{\bd z}} && \bd w^H \bd R_f \bd w + \beta \|\bd w^H \bd A_s - \bd f\|_2^2 + \rho \bd c^T(\bar{\bd z}^{(k)}) \bar{\bd z} \nn\\
\t{s.t.} && \bd w^H \bd a(\theta_0) = 1, \; |\bd w| \leq \bar{\bd z},
\end{eqnarray}
where $\bar{\bd z} \in \{0,1\}^N$ is the antenna selection vector and $\bd c(\bar{\bd z}^{(k)}) = \bd 1_N \oslash (\bar{\bd z}^{(k)} + \gamma)$ with $\gamma$ a small value for preventing explosion. Though formulation in Eq. (\ref{eq:expre_reweight}) is robust against the choice of initial search point by initializing $\bd c=\bd 1_N$, it fails to control the cardinality of the selection vector. The detailed implementation procedure of regularized adaptive sparse array (RASA) design is presented in Algorithm \ref{alg:alg_2}. When the iterative minimization of Eq. (\ref{eq:expre_reweight}) converges, the obtained selection vector $\bar{\bd z}^{(k+1)}$ can be used as an initial point of Eq. (33). Though $\bar{\bd z}^{(k+1)}$ does not satisfy the regularized antenna positions, the iterative optimization of Eq. (\ref{eq:expre_adaptives}) will still converge as proved by Proposition 1.
\begin{algorithm}[!ht]
  \caption{Regularized Adaptive Sparse Array (RASA) Design}
  \label{alg:alg_2}
  \SetKwInOut{Input}{Input} \SetKwInOut{Output}{Output}
  \Input{A set of complementary sparse arrays $\bd Z$ designed by DCSA and parameter $\beta$.}
  \Output{Optimum adaptive sparse array $\bd z$.}
  \For{$m=1, \ldots, M$}{
  \begin{itemize}
  \item Switch on the $m$th quiescent sparse array;
  \item Implement spatial filtering per Eq. (\ref{eq:expre_weight11});
  \item Collect $T$ samples and store in the matrix $\bd Y$;
\
  \end{itemize}
   }
   Calculate the covariance $\bd R_f=(1/T)\bd Y \bd Y^H$;\\
   Set iteration number $k = 0$ and initialize $\bar{\bd z}^{(k)}= \bd 1$;\\
   \Repeat{ $\|\bar{\bd z}^{(k+1)}-\bar{\bd z}^{(k)}\|_2$ is sufficiently small }{
   Calculate $\bd c(\bar{\bd z}^{(k)}) = \bd 1_N \oslash (\bar{\bd z}^{(k)} + \gamma)$;\\
   Solve Eq. (\ref{eq:expre_reweight}) to get $\bar{\bd z}^{(k+1)}$ and $k=k+1$;\\  
   } 
   Set iteration number $k = 0$ and $\bd z^{(k)} = \bar{\bd z}^{(k+1)}$;\\
   \Repeat{ $\|\bd z^{(k+1)}-\bd z^{(k)}\|_2$ is sufficiently small }{
   Update $\bm{\tau}$ according to
   \begin{equation}
   \tau_n = \begin{cases}
   z^{(k)}_n - \zeta, \; \text{if} \; z^{(k)}_n \geq \kappa,\\
   z^{(k)}_n + \zeta, \; \text{if} \; z^{(k)}_n <\kappa,
   \end{cases}
   \end{equation}
   Calculate $\bd g(\bd z^{(k)},\bm{\tau})$ per Eq. (19);\\
   Solve Eq. (\ref{eq:expre_adaptives}) to get $\bd z^{(k+1)}$ and $k=k+1$;\\  
   } 
\end{algorithm}

\section{Simulations}

\label{sec:simulations}

Extensive simulation results are presented in this section to validate the proposed strategy for adaptive sparse array beamformer design. We first demonstrate the effectiveness of proposed DCSA algorithm for solving the cardinality-constrained optimization problem, followed by RCAS strategy validation in both static and dynamic environment. 

\subsection{Algorithm Validation}

\begin{figure}
	\centering
	\includegraphics[scale=0.55]{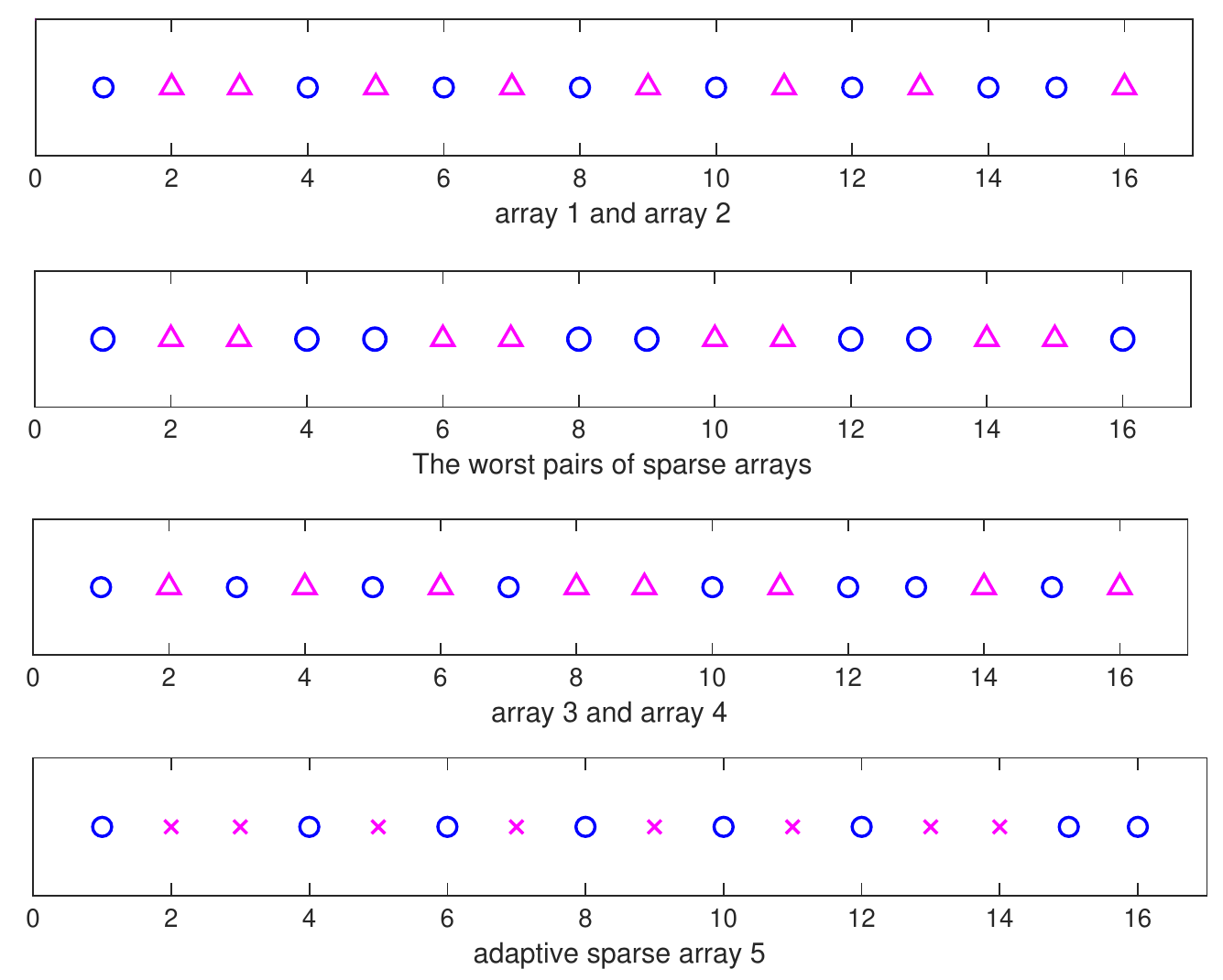}
	\caption{The obtained optimum and worst array splitting by enumeration (upper plot) and Algorithm 1 (middle plot). The marker ``circle'' denotes one array and ``triangle'' denotes the other. The bottom plot is the adaptive sparse array, where ``circle'' denotes selected and ``cross'' denotes discarded.}
	\label{fig_splitarray}
\end{figure}

\begin{figure}
	\centering
	\includegraphics[scale=0.55]{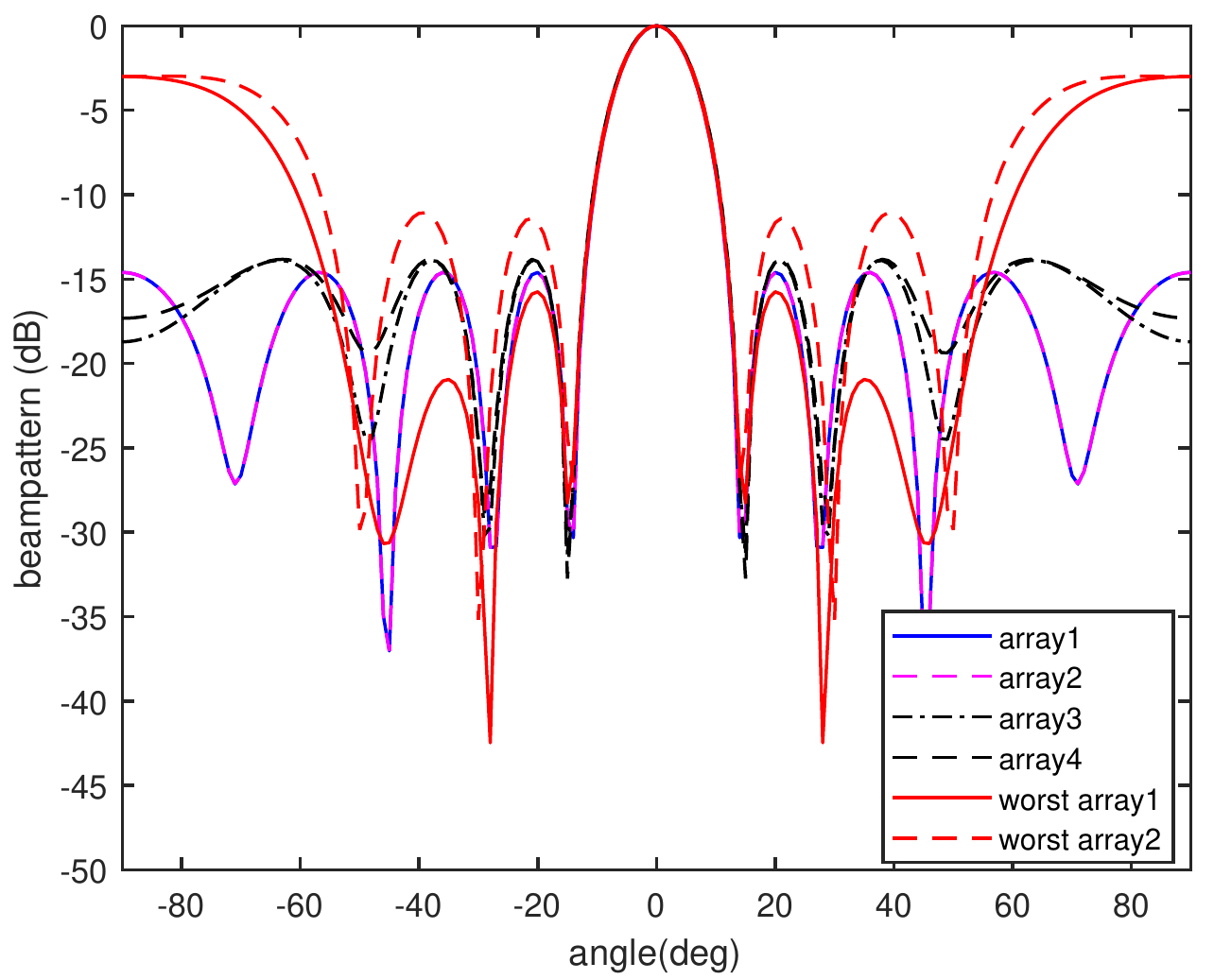}
	\caption{The quiescent beampatterns of four sparse arrays 1-4 shown in Fig. \ref{fig_splitarray} and the worst pair of sparse arrays.}
	\label{fig_bp}
\end{figure}

First, we utilize a small array to validate the effectiveness of our proposed algorithm for solving the cardinality-constrained optimization problem. Assume a uniform linear array (ULA) comprising 16 antennas with an inter-element spacing of a quarter wavelength. This linear array is divided into 8 groups and each group contains two consecutive antennas. There are totally 8 front-ends installed and each front-end is responsible to connect with the two adjacent antennas in one group. We split this linear array into two sparse arrays through regularized antenna switching and each sparse array comprises 8 elements. The array is steered towards the broadside direction and the sidelobe angular region is defined as $[-90^{\circ},-12^{\circ}] \cup [12^{\circ},90^{\circ}]$ and the desired sidelobe level is $-15$dB. It is desirable that both split sparse arrays exhibit good quiescent beampatterns. As there are 128 different splitting (resulting in 128 pairs of complementary sparse arrays) in total, we enumerate the peak sidelobe levels (PSLs) of each pair among these 128 pairs for the lowest one. The optimum pair of sparse arrays, named as array 1 and 2, are shown in the upper plot of Fig. \ref{fig_splitarray}, followed by the worst pair of splitting. Clearly, arrays 1 and 2 are the same except of a reversal antenna placement. We then run Algorithm 1, and the finally converged splitting is shown in the lower plot of Fig. \ref{fig_splitarray}. The quiescent beampatterns of these four sparse arrays are depicted in Fig. \ref{fig_bp}, where the worst pair of splitting is plotted as well. We can see that the PSL of the first pair returned by enumeration is -15dB, while that of the second pair returned by Algorithm 1 is slightly larger than -15dB. Although inferior to the global optimum splitting, the proposed cardinality-constrained optimization algorithm could seek a sub-optimal solution with an acceptable performance.


\begin{figure}
	\centering
	\includegraphics[scale=0.55]{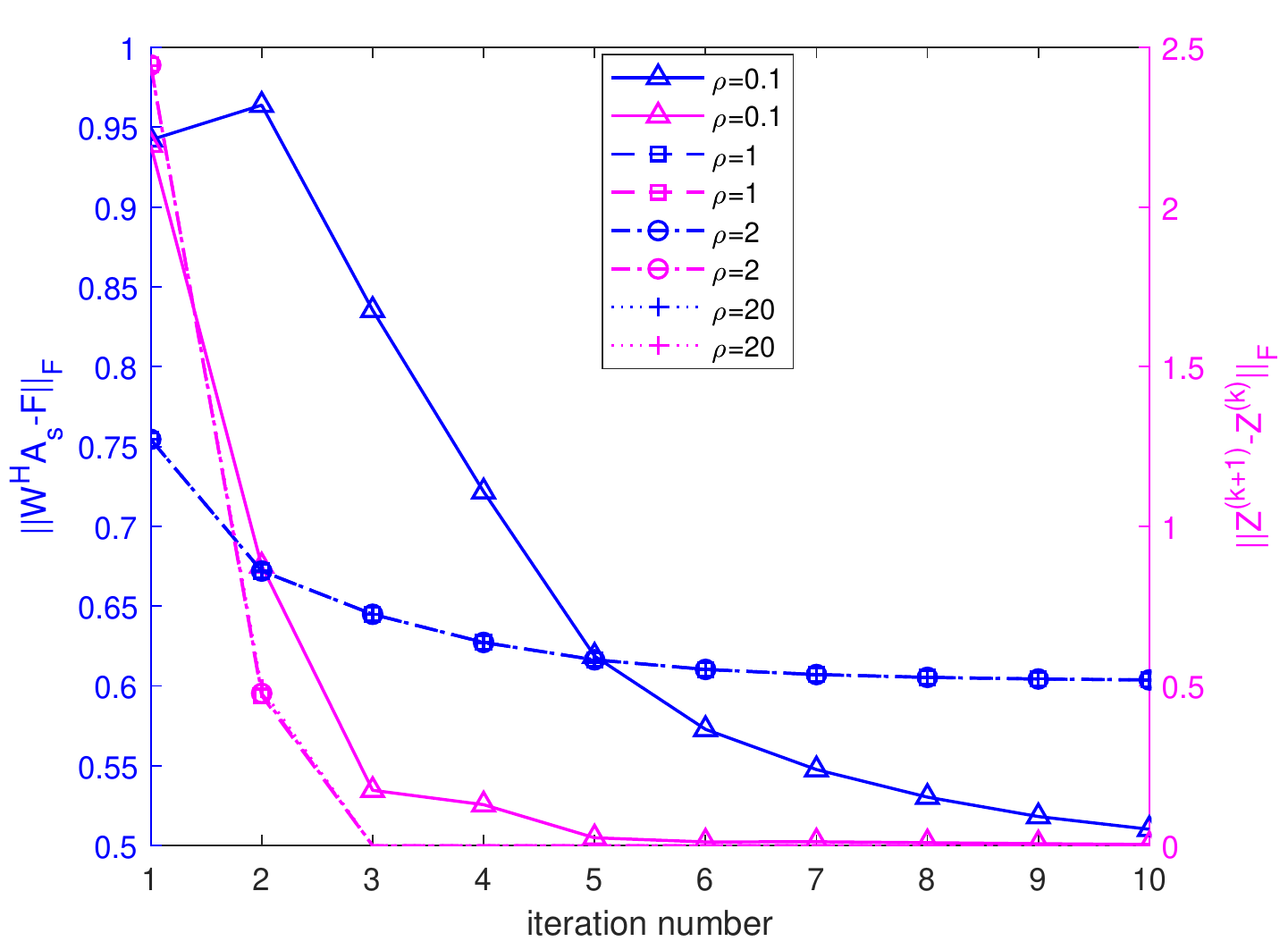}
	\caption{Convergence performance: $\|\bd W^H \bd A_s-\bd F\|_{\t{F}}$ and $\|\bd Z^{(k+1)}-\bd Z^{(k)}\|_{\t{F}}$ versus iteration number in four cases of different $\rho$.}
	\label{fig_algconverge}
\end{figure}

We continue to investigate the convergence performance of the proposed algorithm. The trade-off parameter in ($RP_{\tau}$) is first set as $\rho=1$ in this considered example. We plot two parts of the objective function in ($RP_{\tau}$), those are $\|\bd W^H \bd A_s-\bd F\|_{\t{F}}$ and $\|\bd Z^{(k+1)}-\bd Z^{(k)}\|_{\t{F}}$, versus the iteration number. The results are plotted in Fig. \ref{fig_algconverge}. We can see that the auxiliary variable $\bd Z$ requires maximally three iterations to converge, while the beamforming weight matrix $\bd W$ requires more iterations. The reason is explained as that the iterative updating of the beampattern phase $\bd F_p$ will further decrease the beampattern deviation after the sparse array is designed. We then change the value of the trade-off parameter $\rho$ discretely to four different values of $0.1, 1, 2, 20$. The curves of two parts of the objective function versus the iteration number are plotted in Fig. \ref{fig_algconverge} as well. We can see that when $\rho \geq 1$, the effect of $\rho$ on the convergence rate and synthesized beampattern shape is diminished and negligible.

%
%


\subsection{Static Environment}

\begin{figure*}[t!]
\begin{minipage}{0.23\linewidth}
  \centerline{\includegraphics[scale=0.35]{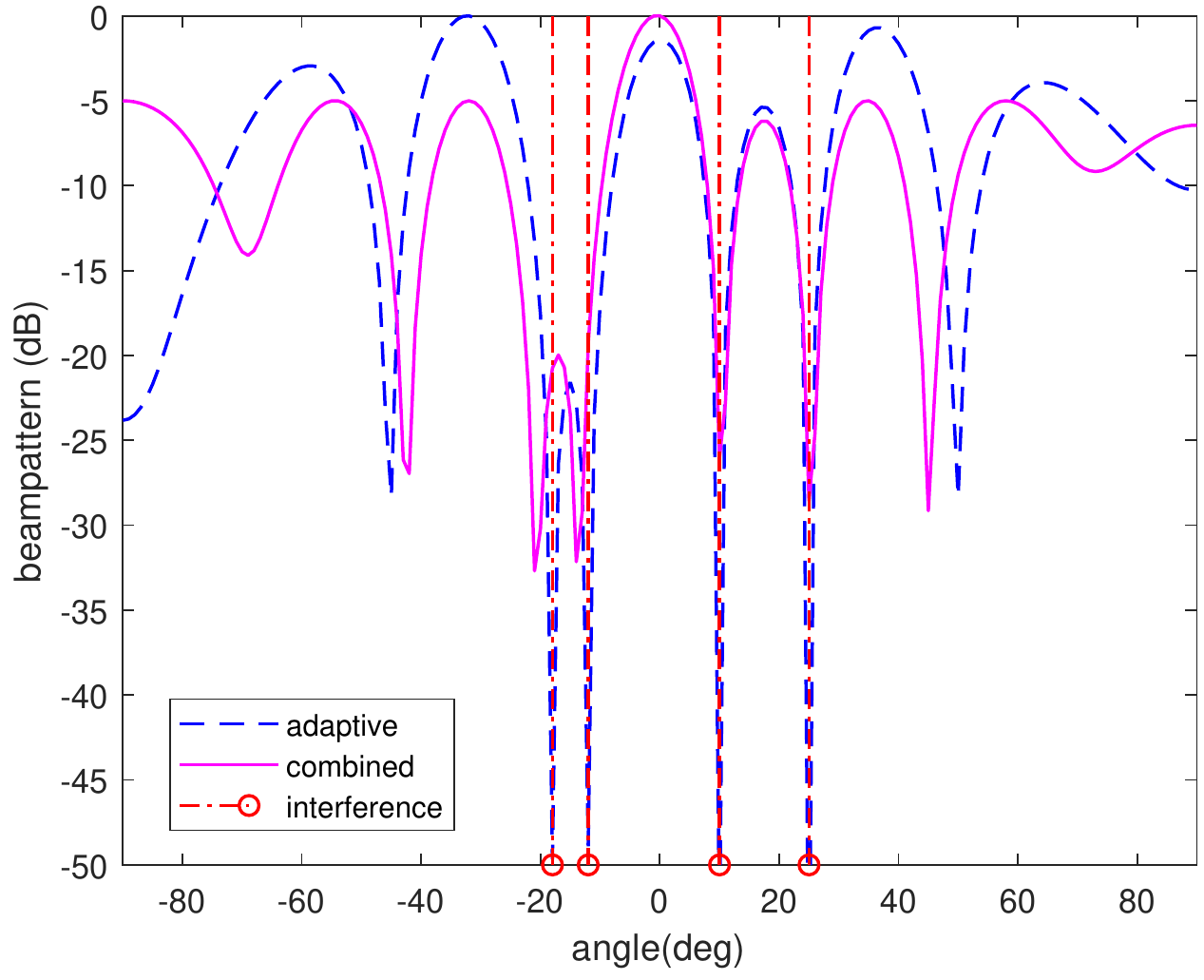}}
  \centerline{(a)}
\end{minipage}
\hfill
\begin{minipage}{0.23\linewidth}
  \centerline{\includegraphics[scale=0.35]{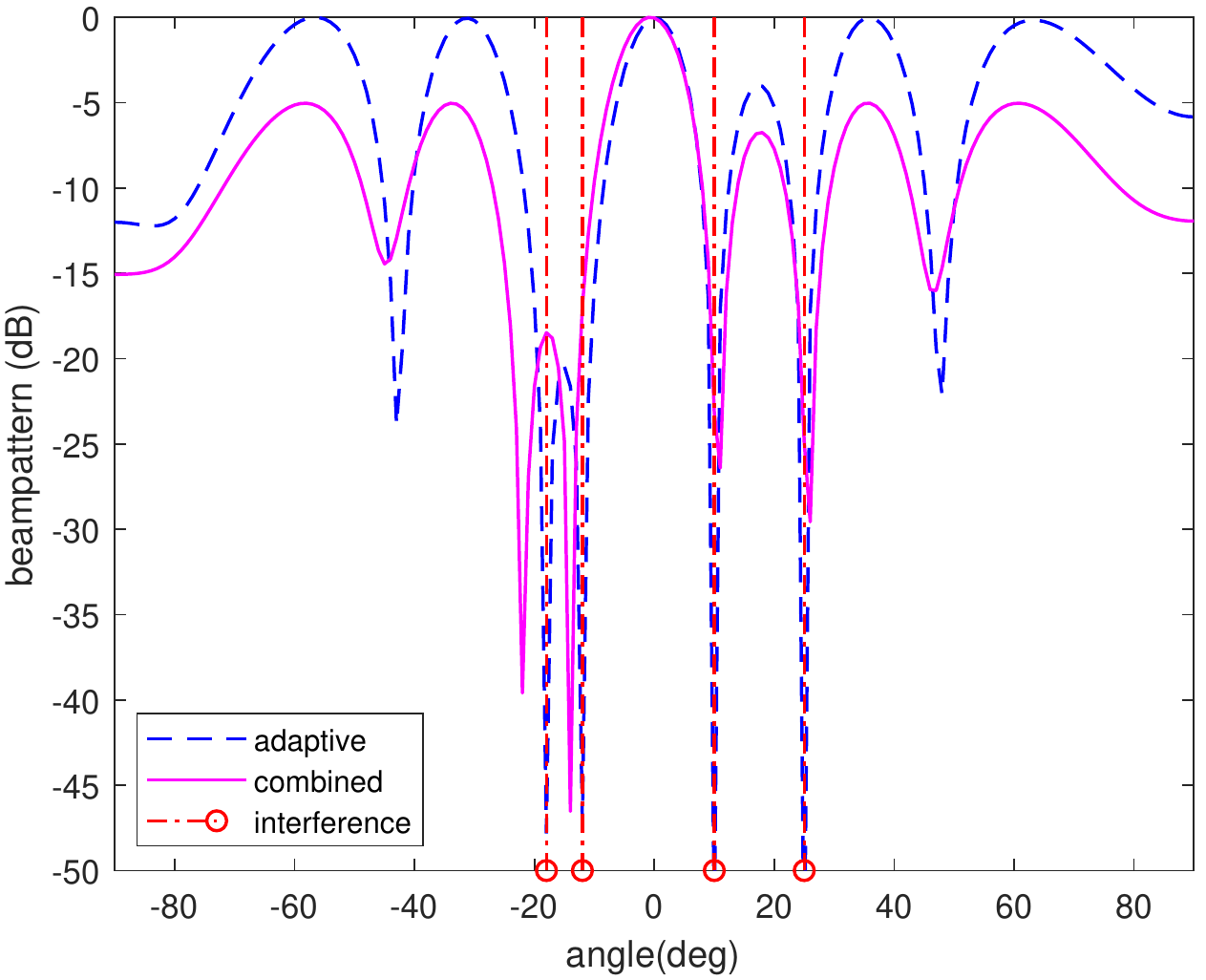}}
  \centerline{(b)}
\end{minipage}
\hfill
\begin{minipage}{0.23\linewidth}
  \centerline{\includegraphics[scale=0.35]{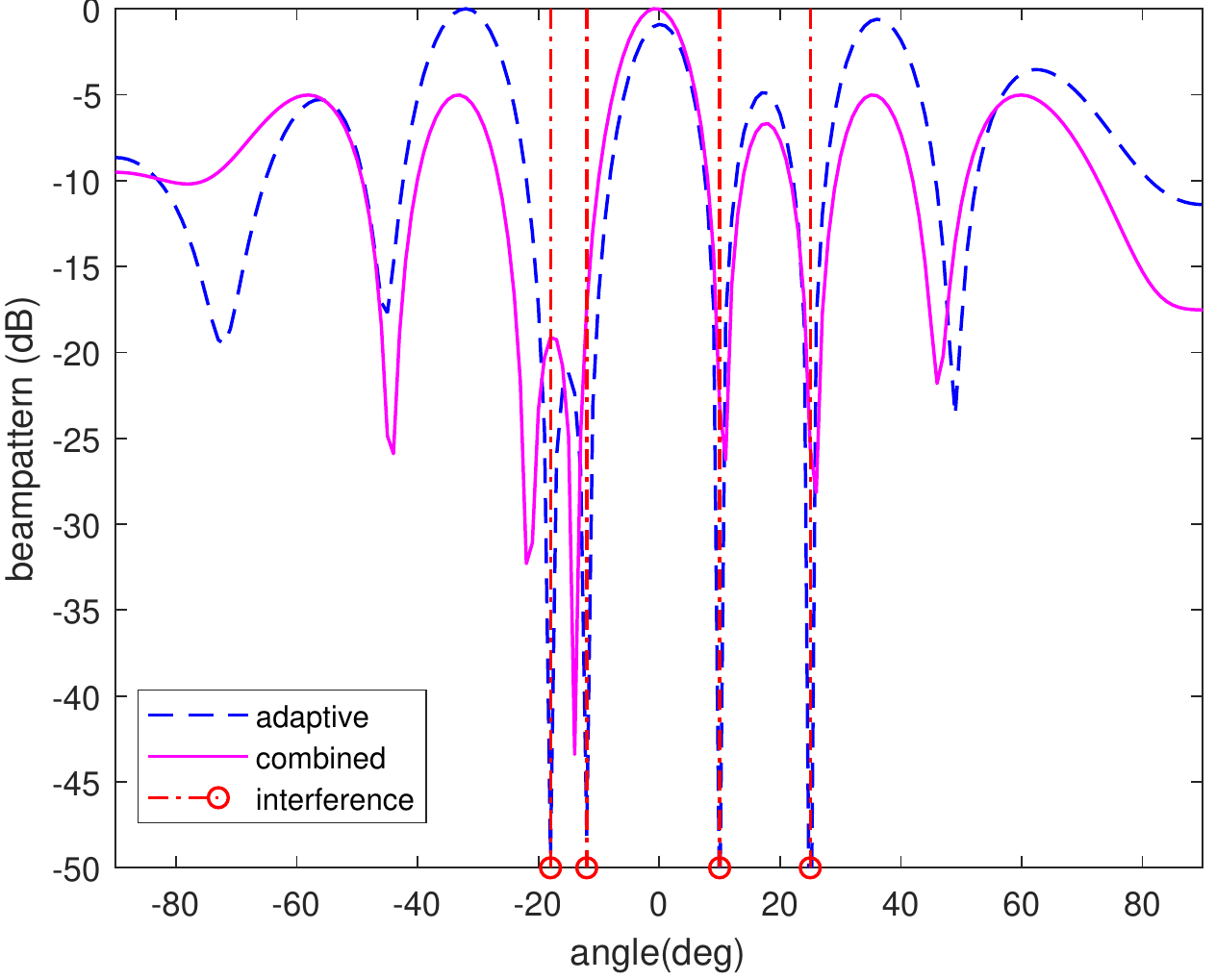}}
  \centerline{(c)}
\end{minipage}
\hfill
\begin{minipage}{0.23\linewidth}
  \centerline{\includegraphics[scale=0.35]{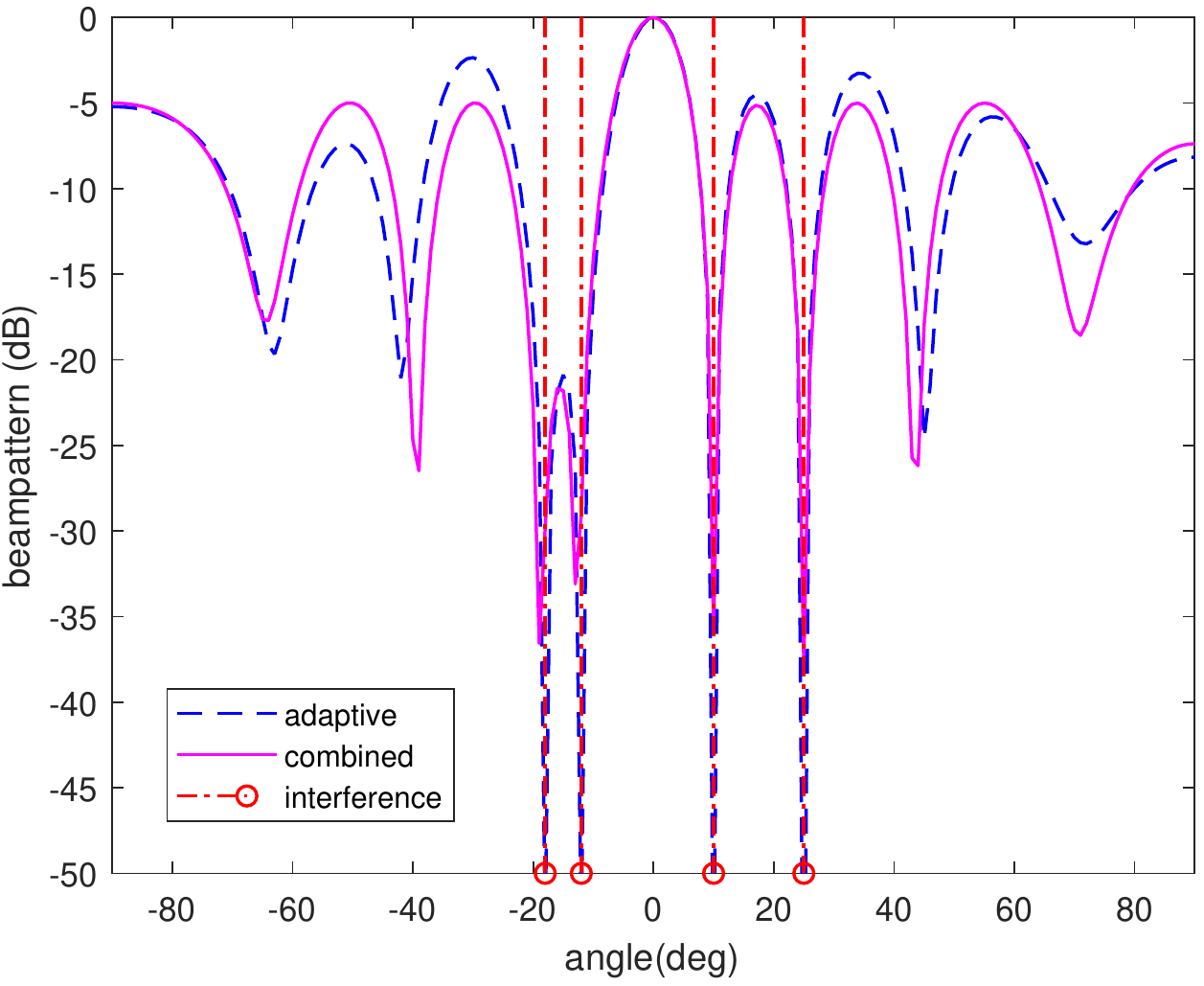}}
  \centerline{(d)}
\end{minipage}
\caption{The beampatterns of three sparse arrays (1)-(5) using both adaptive beamforming and combined beamforming: (a) Array 1 and 2 (b) Array 3 (c) Array 4 (d) Array 5.}
\label{fig_bpCSAS1}
\end{figure*}

As explained in our previous work \cite{Wang2014}, array configuration plays a very sensitive and viral role in the scenario of closely-spaced mainlobe interferences, whereby the importance of sparse arrays is amplified. Hence, we intentionally consider such cases in this work. Assume that four interferences are coming from the directions of $-28^{\circ},-12^{\circ},10^{\circ},25^{\circ}$ with an interference-to-noise ratio (INR) of 20dB. The proposed RCAS is conducted, where two quiescent sparse arrays 3 and 4 are first sequentially switched on to collect the data of the full array, and then the optimum adaptive sparse array is configured according to Algorithm 2 with the desired SLL setting as $-5$dB. The structure of the adaptive sparse array is shown in the bottom plot of Fig. \ref{fig_splitarray}. The beampatterns of three sparse arrays (3)-(5) using both adaptive beamforming in Eq. (\ref{eq:expre_weightcapon}) and combined beamforming in Eq. (\ref{eq:expre_weight11}) are presented in Fig. \ref{fig_bpCSAS1}. We can observe that (1) Adaptive beamforming, though excellent in suppressing interferences, exhibits catastrophic grating lobes except for the sparse array 5. This inadvertently manifests the superiority of the designed sparse array 5, irrespective of no explicit sidelobe constraints imposed in adaptive beamforming. (2) Combined beamforming is capable of controlling the sidelobe level with the sacrifice of shallow nulls towards the unwanted interferences; (3) Arrays 3 and 4 with combined beamforming cannot form nulls towards the two closely-spaced interferences from $-18^{\circ}$ and $-12^{\circ}$. The output SINR of five sparse arrays are listed in Table \ref{table_1}. We can obtain the corresponding results with Fig. \ref{fig_bpCSAS1}, that is array 5 is supreme in terms of the output SINR regardless of beamforming methods, while arrays 1(2) and 3(4) exhibit much worse performance especially for combined beamforming.

\begin{table}
\center
\caption{Output SINR of five sparse arrays in Fig. \ref{fig_splitarray}.}
\begin{threeparttable}
\begin{tabular}{|c|c|c|c|c|c|}
\hline
\hline
\backslashbox{B}{A} & 1 & 2 & 3 & 4 & 5 \\
\hline 
Adaptive & 4.28dB & 4.28dB & 3.84dB & 4.28dB & 6.52dB\\
\hline
Combined & -4.44dB & -4.44dB & -6.58dB & -6.2dB & 3.92dB \\
\hline
\hline
\end{tabular}
    \begin{tablenotes}
      \small
      \item ``B'' denotes beamforming method and ``A'' denotes array numbering. The unit is in dB.
    \end{tablenotes}
  \end{threeparttable}
\label{table_1}
\end{table}

\begin{figure}
	\centering
	\includegraphics[scale=0.55]{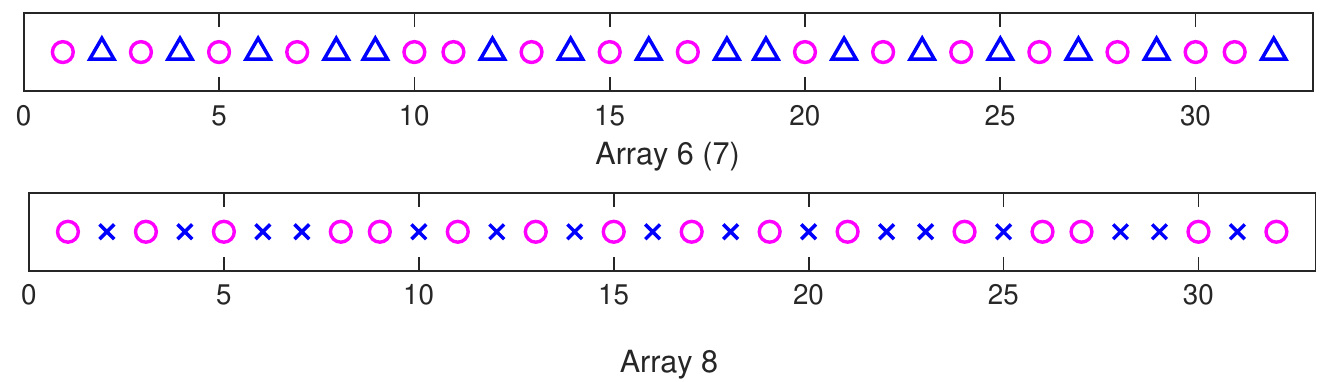}
	\caption{Configurations of sparse arrays 6-8: arrays 6 and 7 are quiescent complementary and array 8 is adaptive.}
	\label{fig_arraylarge}
\end{figure}

\begin{figure}
	\centering
	\includegraphics[scale=0.55]{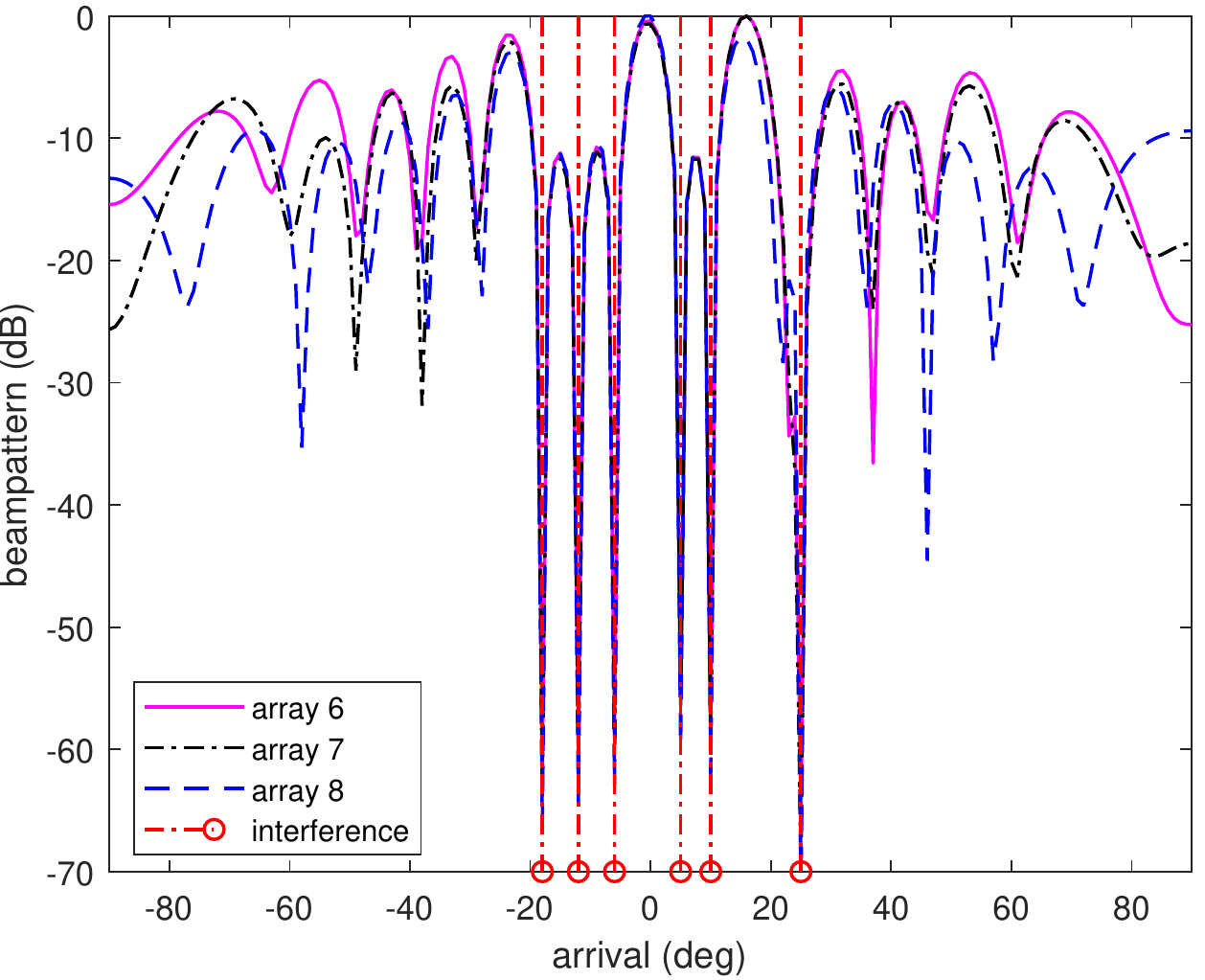}
	\caption{Adaptive beamforming of sparse arrays 6, 7 and 8.}
	\label{fig_bp_largeq}
\end{figure}

\begin{figure}
	\centering
	\includegraphics[scale=0.55]{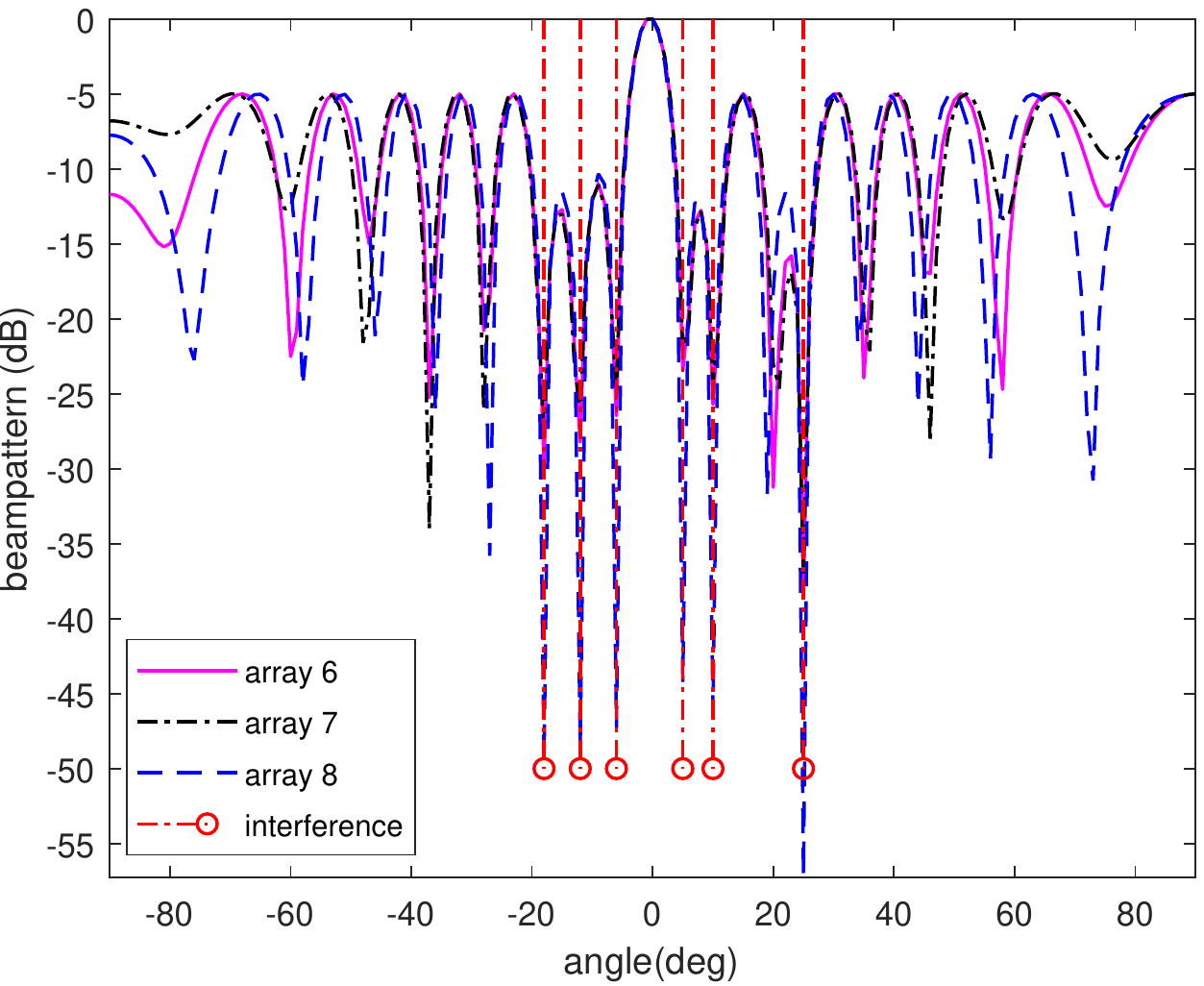}
	\caption{Combined beamforming of sparse arrays 6, 7 and 8.}
	\label{fig_bp_largec}
\end{figure}

To thoroughly examine the performance of the proposed RCAS strategy, we proceed to enlarge the array size to a ULA with 32 antennas. This linear array is divided into 16 groups and each group contains two consecutive antennas. In the first step of RCAS, we design a set of complementary sparse arrays, which are obtained by the extended Algorithm 1 and shown in the upper plot of Fig. \ref{fig_arraylarge}. The sidelobe angular region is defined as $[-90^{\circ},-7^{\circ}] \cup [7^{\circ},90^{\circ}]$ and the desired sidelobe level is set as $-5$dB for combined beamforming. Six interferences are impinging on the array from $-18^{\circ},-12^{\circ},-6^{\circ},5^{\circ},10^{\circ},25^{\circ}$ with an INR of 20dB. The configuration of adaptive sparse array is shown in the lower plot of Fig. \ref{fig_arraylarge}. The adaptive and combined beampatterns of three arrays 6-8 are compared in Fig. \ref{fig_bp_largeq} and \ref{fig_bp_largec}, respectively. Again, adaptive beamforming avoidably produces very high sidelobes (even grating lobes in a severe condition), which adversely affect the output performance in the dynamic environment where an unintentional interference is suddenly switched on. We can see that array 8 produces the deepest nulls towards the interferences, in turn yielding the highest SINR. Comparatively, the output SINR of sparse arrays 6-8 are listed in Table \ref{table_2} for both adaptive and combined beamforming.

\begin{table}
\center
\caption{Output SINR of three sparse arrays in Fig. \ref{fig_arraylarge}.}
\begin{tabular}{|c|c|c|c|}
\hline
\hline
\backslashbox{Beamforming}{Array} & 6 & 7 & 8 \\
\hline 
Adaptive & 6.8dB & 6.99dB & 8.44dB \\
\hline
Combined & -1.78dB & -2.95dB & 7.44dB \\
\hline
\hline
\end{tabular}
\label{table_2}
\end{table}

\subsection{Dynamic Environment}

\begin{figure}
	\centering
	\includegraphics[scale=0.55]{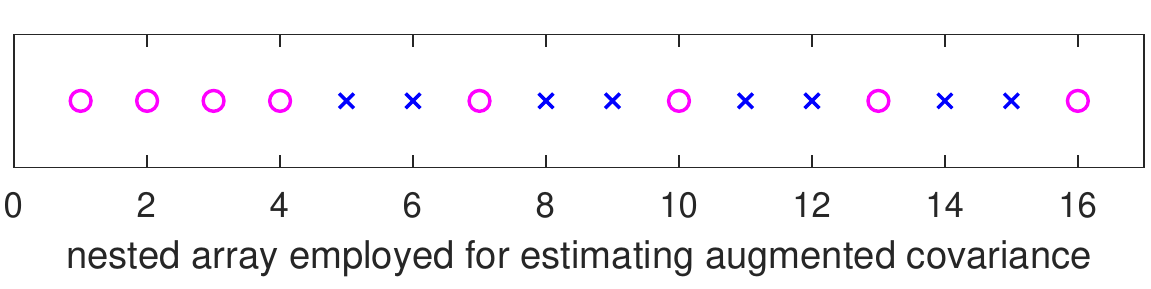}
	\caption{The nested array deliberately used for augmented full covariance estimation.}
	\label{fig_arraynest}
\end{figure}

Different from the proposed RCAS strategy, the environmental information is obtained from the augmented covariance matrix of a virtual full co-array\cite{Hamza2019,Pal2010,Hamza2020} and is referred to as coarray strategy. To give a brief explanation, the covariance matrix $\bd R$ of the physical array is first vectorized and then reordered according to the spatial lags. Both spatial smoothing and Toepliz matrix augmentation  can then be utilized to obtain the full covariance matrix of the virtual co-array. It would be intriguing to compare the difference of array configuration and filtering performance based on the knowledge acquired from the two strategies. As it is impossible to construct a fully-augmentable sparse array under the constraint of regularized antenna switching, a nested array in Fig. \ref{fig_arraynest} is deliberately employed for situational sensing and performance comparison.

\begin{figure}
	\centering
	\includegraphics[scale=0.55]{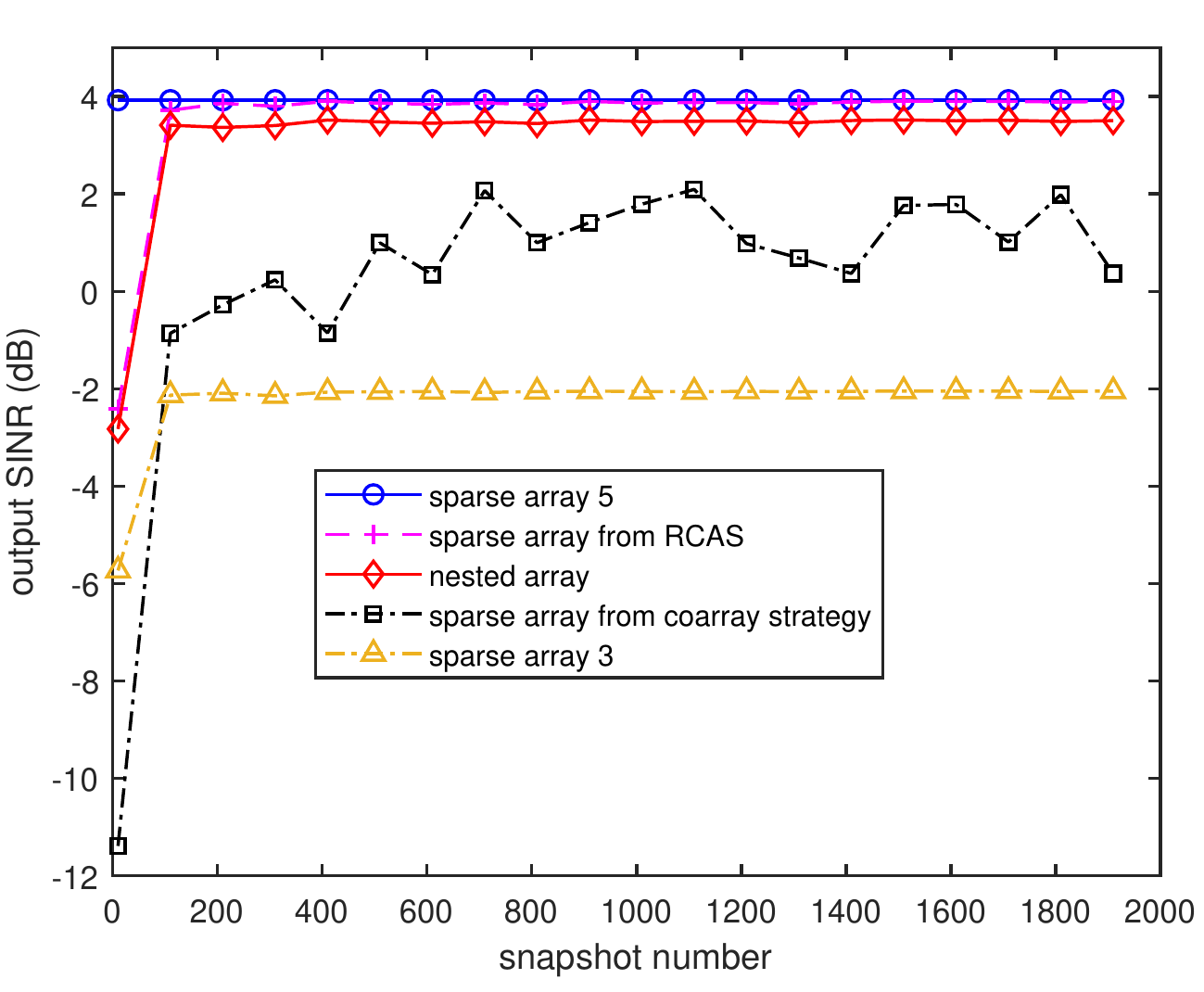}
	\caption{Output SINR versus snapshot number in the case of uncorrelated interfering signals.}
	\label{fig_sinrsnapshotun}
\end{figure}

\begin{figure}
	\centering
	\includegraphics[scale=0.55]{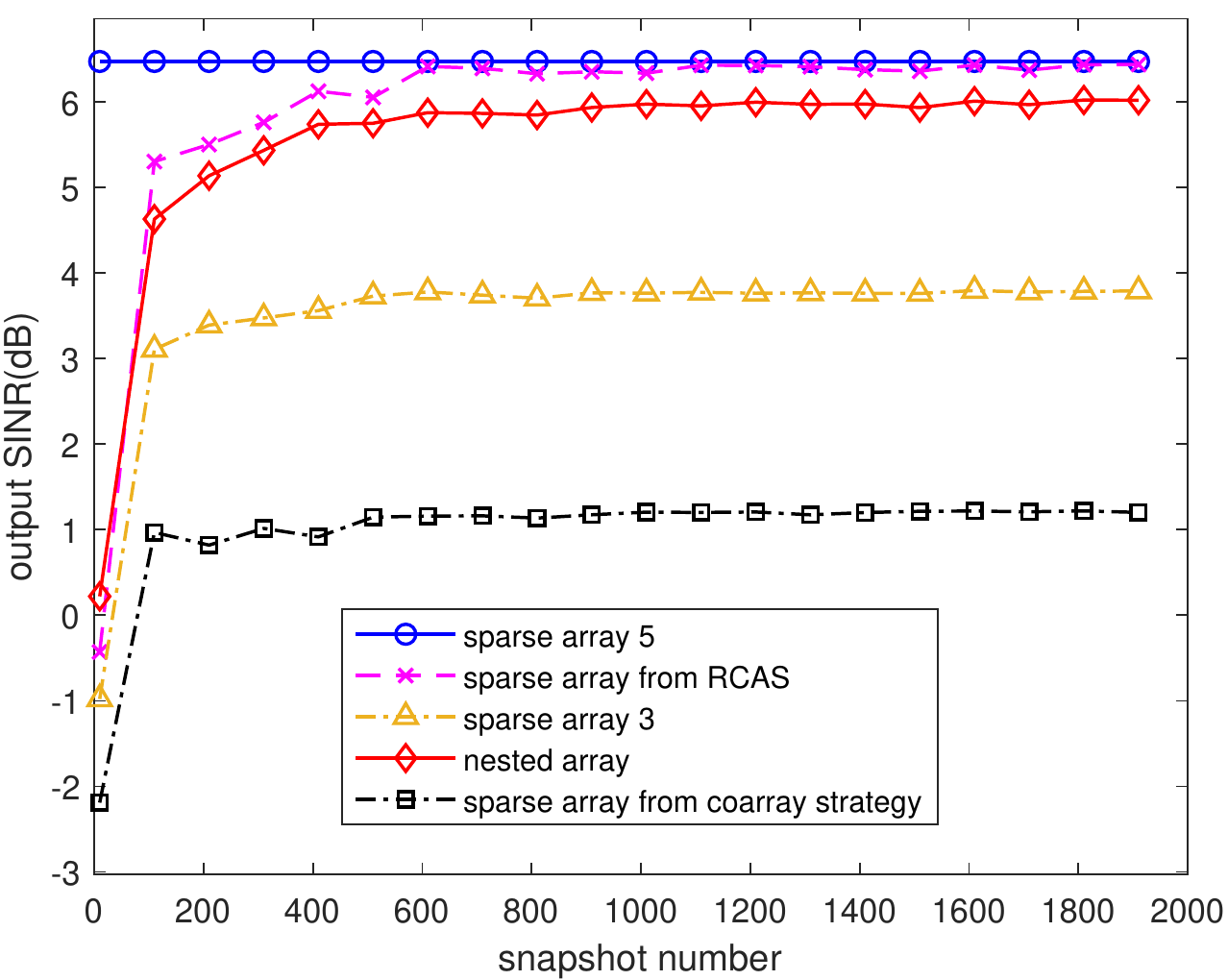}
	\caption{Output SINR versus snapshot number in the case of correlated interfering signals.}
	\label{fig_sinrsnapshotco}
\end{figure}

We first examine the effect of snapshot number on both strategies. The simulation scenario remains the same as that of the above small array. We choose the sparse arrays 3 and 5 as the benchmark and compare the output SINR of optimized sparse arrays using two strategies. The snapshot number is changing from 10 to 1910 in a step of 100 and 1000 Monte Carlo simulations are run in each case. For each number, we collect the data of corresponding length either switching between sparse arrays 3 and 4 or using the nested array and augmenting the covariance. For more details on covariance augmentation, the readers can refer to \cite{Pal2010,Hamza2019} and reference therein. It is worth noting that the estimation accuracy of the full array covariance matrix significantly affect the selection of switched antennas. Hence, for different numbers of snapshots, Algorithm 2 is utilized to calculate the optimum adaptive sparse array with the input of estimated full array covariance. The curves of output SINR versus snapshot number in two cases of uncorrelated and correlated interfering signals are depicted in Fig. \ref{fig_sinrsnapshotun} and \ref{fig_sinrsnapshotco}, respectively. For the latter, the correlation among different interferences are generated randomly. The proposed RCAS strategy can return the optimum sparse array that is array 5 when the snapshot number increases. Although the configuration of nested array is not restricted to regularized antenna positions, its output SINR does not exhibit superiority over sparse array 5. The sparse array configured based on the environmental information obtained from augmented covariance matrix does not exhibit satisfactory performance especially when the snapshot number is small and the impinging interferences are correlated. This attributes to the fact that coarray-based signal processing usually requires a large number of snapshots and is restricted to dealing with uncorrelated signals. 

%

\begin{figure}
	\centering
	\includegraphics[scale=0.45]{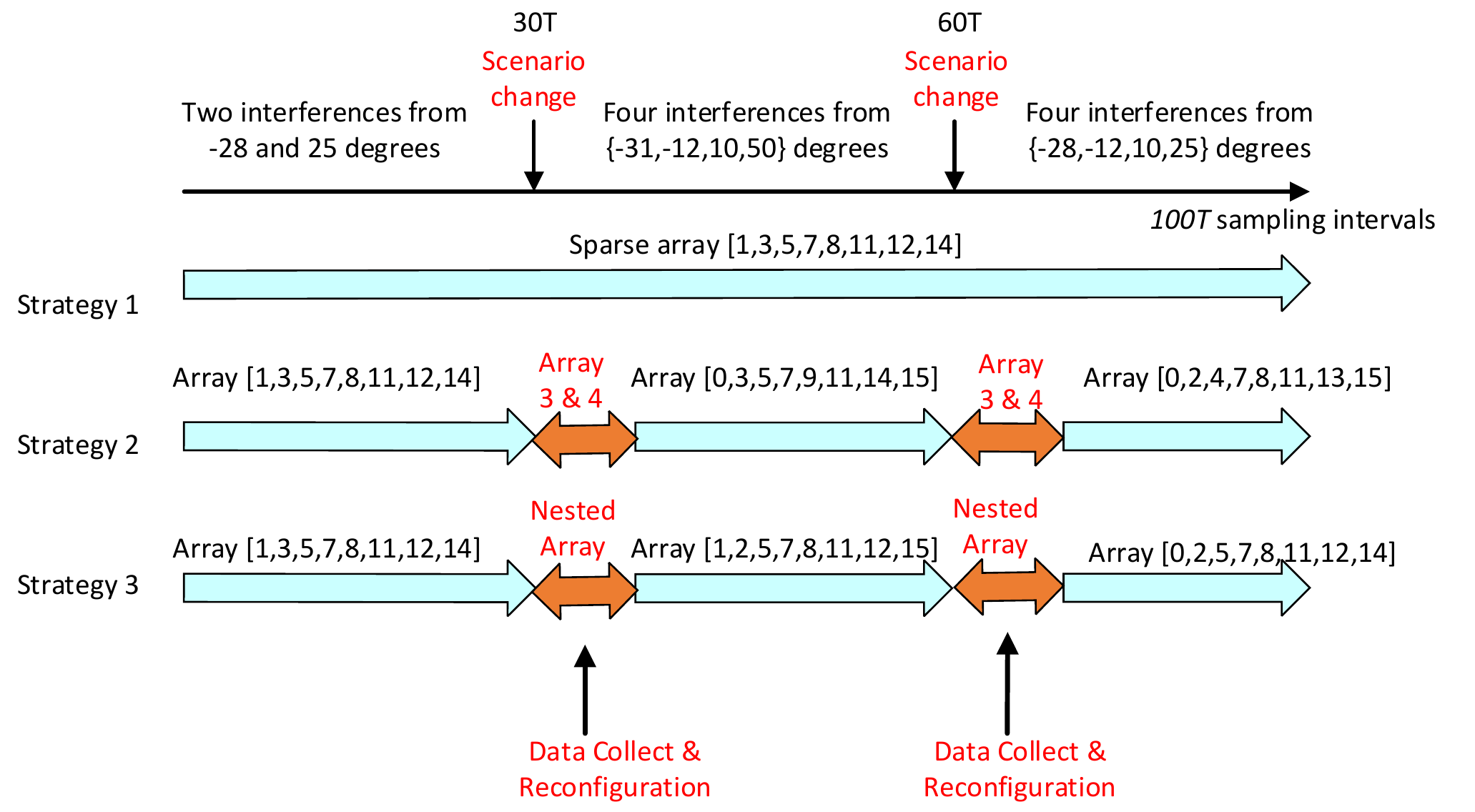}
	\caption{Three strategies in dynamic environment and the antenna positions of arrays utilized in each strategy are indicated.}
	\label{fig_sinrscenario}
\end{figure}

\begin{figure}
	\centering
	\includegraphics[scale=0.6]{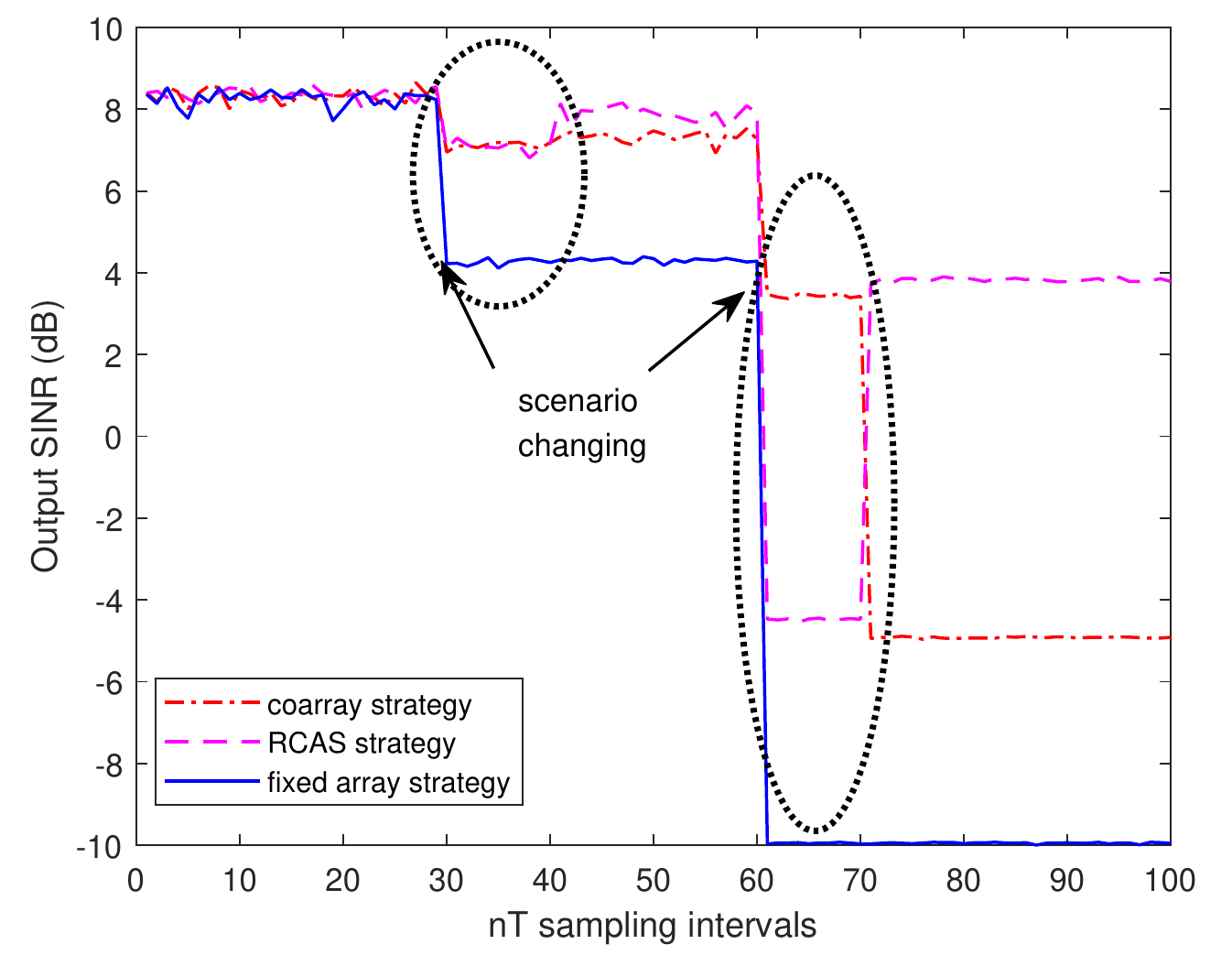}
	\caption{Comparison of three strategies in dynamic environment. The environment has changed twice during the observation time as indicated by black arrows. The periods highlighted by black circles are stages for environmental sensing, data collection and array reconfiguration.}
	\label{fig_sinrscenario1}
\end{figure}

We continue to examine an example of dynamic environment, which is described in Fig. \ref{fig_sinrscenario}. The total observing time is set as $100T$ sampling intervals with $T=500$. Suppose that the target is coming from broadside, and there are two interferences coming from $-28^{\circ}$ and $25^{\circ}$ in the first period of the observing time. At the time instant of $30$T, the scenario has changed to that of four interferences coming from $-31^{\circ}$, $-12^{\circ}$, $10^{\circ}$ and $50^{\circ}$. At the time instant of $60$T, the scenario has changed again and four interferences are coming from $-28^{\circ}$, $-12^{\circ}$, $10^{\circ}$ and $25^{\circ}$, respectively. The INR of all interferences is 20dB. We consider three strategies, those are fixed-array strategy, the proposed RCAS strategy and coarray strategy. In the first strategy, a fixed sparse array, that is the optimal sparse array configured for combined beamforming in scenario 1 using Eq. (\ref{eq:expre_reweight}), is employed during the observing period. In the proposed RCAS strategy, complementary sparse arrays 3 and 4 are utilized for environment sensing and data collection, and adaptive sparse arrays are then configured based on the full array covariance, as indicated in the middle row of Fig. \ref{fig_sinrscenario}. In the coarray strategy, the full array covariance is obtained by augmenting the spatial lags of the nested array shown in Fig. \ref{fig_arraynest}, according to the structure of Toeplitz matrix. The two adaptive sparse arrays are optimized based on the virtual covariance, as indicated in the bottom row of Fig. \ref{fig_sinrscenario}. We can see from Fig. \ref{fig_sinrscenario1} that the fixed sparse array experiences performance degradation at two durations of scenario changing. Though the nested array performs better than complementary sparse arrays during the second sensing stage, its configuration does not follow the rule of regularized antenna switching. The proposed RCAS strategy exhibits the best performance after adaptive sparse array reconfiguration and modest output SINR during the two environment sensing stages. 

\section{Conclusions}
\label{sec:conclusion}

A complementary sparse array switching (RCAS) strategy was proposed in this work for adaptive sparse array beamformer design, which aimed to swiftly adapt intertwined array configuration and excitation weights in accordance to dynamic environment. Our previous work usually assumed either known or estimated environmental information, which was regarded as an impediment to practical implementation. The RCAS works in two steps. First, a set of deterministic complementary sparse arrays, all with good quiescent beampatterns, was designed and a full data collection was conducted by switching among them. Then, the adaptive sparse array was configured for the specific environment, based on the data collected in the first step. Deterministic and adaptive sparse arrays in both design steps were restricted to regularized antenna switching for increased practicability. The RCAS was devised as an exclusive cardinality-constrained optimization and an iterative algorithm was proposed to solve it effectively. We conducted a rigorous theoretical analysis and proved that the proposed algorithm is an equivalent transformation to the original cardinality-constrained optimization. Extensive simulation results validated the effectiveness of proposed RCAS strategy. It would be desirable that the adaptive sparse array design would eliminate the requirement of either environmental information or full array covariance. The solution to this problem will be investigated in the near future.

\section{Appendix}

\subsection{Proof of Lemma 1}
\label{subsec:lemma1}

(1) The cardinality of a vector $\bd b \in \mathbb{R}_+^M$ is expressed as,
\begin{equation}
\label{eq:expre_norm0}
\|\bd b\|_0 = \bd 1_M^T \t{sign}(\bd b),
\end{equation}
where the sign function is defined as
\begin{equation}
\label{eq:expre_sign}
\t{sign}(\cdot) = \begin{cases}
1, & \t{if} \; \cdot > 0 \\
0. & \t{if} \; \cdot = 0
\end{cases}
\end{equation}
According to the definition of $\phi(\bd b,\bm{\tau})$, it is separable and can be rewritten as
\begin{equation}
\label{eq:expre_phi}
\phi(\bd b,\bm{\tau}) = \sum_{i=1}^M \phi(b_i,\tau_i),
\end{equation}
where $\phi(b_i,\tau_i) = (1/\tau_i)(b_i - (b_i - \tau_i)^+)$ and thus
\begin{equation}
\label{eq:expre_h}
\phi(b_i,\tau_i) = \begin{cases}
1 & \t{if} \; b_i > \tau_i, \\
b_i /\tau_i & \t{if} \; b_i \leq \tau_i,
\end{cases}
\end{equation}
Here, $(b_i - \tau_i)^+ = \max\{b_i-\tau_i, 0\}$. Comparing the sign function in Eq. (\ref{eq:expre_sign}) with the function $\phi(b_i,\tau_i)$ in Eq. (\ref{eq:expre_h}), we obtain that, $\forall \; b_i \geq 0$
\begin{equation}
\label{eq:expre_com}
0 \leq \phi(b_i,\tau_i) \leq \t{sign}(b_i) \leq 1.
\end{equation}
Combining Eqs. (\ref{eq:expre_norm0}), (\ref{eq:expre_phi}) and (\ref{eq:expre_com}) yields $\phi(\bd b,\bm{\tau}) \leq \|\bd b\|_0$, that is $\phi(\bd b,\bm{\tau})$ is a piecewise linear under-estimator of $\|\bd b\|_0$. Moreover, function $\phi(b,\tau)$ is obviously a non-increasing function of $\tau$, that is,
\begin{equation}
\phi(b, \tau_1) \geq \phi(b, \tau_2), \; \t{if} \; \tau_1 \leq \tau_2
\end{equation}
Therefore, for any given vector $\bf b$, we have that $\phi(\bd b,\bm{\tau}_2) \leq \phi(\bd b,\bm{\tau}_1)$ from Eq. (\ref{eq:expre_phi}), given that $\bm{\tau}_1 \leq \bm{\tau}_2$. Here, $\bm{\tau}_1 \leq \bm{\tau}_2$ implies that $\tau_{1,i} \leq \tau_{2,i}, \forall i=1, \ldots, M$. Thus, $\phi(\bd b, \bm{\tau})$ is a non-increasing function of $\bm{\tau}$.

(2) Comparing Eqs. (\ref{eq:expre_sign}) with (\ref{eq:expre_h}), we also have that
\begin{equation}
\lim_{\tau \rightarrow 0^+} \phi(\cdot, \tau) = \t{sign}(\cdot),
\end{equation}
Thus, utilizing Eqs. (\ref{eq:expre_norm0}) and (\ref{eq:expre_phi}), we have that
\begin{equation}
\lim_{\bm{\tau}\rightarrow 0^+} \phi(\bd b, \bm{\tau}) = \|\bd b\|_0.
\end{equation}

(3) For arbitrary two points $\bd b_1$ and $\bd b_2$. we have that $\forall \; 0 \leq \alpha \leq 1$,
\begin{equation}
\alpha \phi(\bd b_1, \bm{\tau}) + (1- \alpha) \phi(\bd b_2, \bm{\tau}) \leq \phi[\alpha \bd b_1 + (1- \alpha) \bd b_2, \bm{\tau}].
\end{equation}
This proves that the piece-wise linear function $\phi(\bd b, \bm{\tau})$ is a concave function with respect to the variable $\bd b$. The definition of sub-gradient is that for any $\bd g(\bd b_0, \bm{\tau})= \frac{\partial \phi(\bd b, \bm{\tau})}{\partial \bd b} \mid_{\bd b_0}$ such that $\forall \; \bd b$, we have the following,
\begin{equation}
\label{eq:expre_defsubgra}
\phi(\bd b,\bm{\tau}) \leq \phi(\bd b_0, \bm{\tau}) + \bd g(\bd b_0, \bm{\tau})^T(\bd b - \bd b_0).
\end{equation}
When $b_i > \tau_i$, the function $\phi(b_i,\tau_i)$ is differentiable and the gradient is zero. When $0 \leq b_i < \tau_i$, the function $\phi(b_i,\tau_i)$ is also differentiable and the gradient is $1/\tau_i$. When $b_i = \tau_i$, the function $\phi(b_i,\tau_i)$ is non-differentiable, and the sub-gradient can be any number between $[0, 1/\tau_i]$ according to Eq. (\ref{eq:expre_defsubgra}).

%
%
%
%

\subsection{Proof of Theorem 1}
\label{subsec:theorem3}

By utilizing the implicit binary constraint of the auxiliary variable $\bd Z \in \{0,1\}^{N \times M}$, the complementary sparse array design in Eq. (\ref{eq:expre_splittinga}) can be rewritten as follows,
\begin{eqnarray}
\label{eq:expre_splitting3_1}
(AP_{b}) \; \min_{\bd W, \bd Z} && \| \bd W^H \bd A_s - \bd F \|_{\t{F}}  \\
\t{s.t.} && \bd W^H \bd a = \bd 1_M, \nn\\
         && |\bd W| \leq \bd Z, \nn \\      
         && \bd Z \in \{0,1\}^{N\times M}, \nn\\
         && \bd 1^T_M \bd P_l \bd Z \bd c_m =1, l=1,\ldots,L, m=1,\ldots,M \nn\\      
         && \bd Z \bd 1_M  = \bd 1_{N}. \nn
\end{eqnarray}
By comparing the two problems in Eqs. (\ref{eq:expre_splitting3}) and (\ref{eq:expre_splitting3_1}), we can observe that though the third constraint exhibits difference, the two formulations are essentially the same. 

%
%

Let us define a set $\underline{\Omega}=\{\bd Z \geq 0: \bd Z \bd 1_M  = \bd 1_{N}, \bd 1^T_M \bd P_l \bd Z \bd c_m =1,l=1,\ldots,L, m=1,\ldots,M\}$, then the domain of problem $(AP_{\tau})$ can be expressed as $\hat{\Omega}_{\tau} = \{\bd Z: \bd Z \in \underline{\Omega}, \phi(\bd P_l \bd Z \bd c_m, \bd P_l \bm{\Pi} \bd c_m) \leq 1, l=1,\ldots,L, m=1,\ldots,M\}$. Similarly, the domain of the problem $(AP_{b})$ is same as that of $(AP_{0})$, that is $\hat{\Omega}_{0} = \{\bd Z: \bd Z \in \underline{\Omega}, \bd Z \in \{0,1\}^{N\times M}\}$. Then, we are going to prove that $\hat{\Omega}_{\tau}=\hat{\Omega}_{0}$.

Suppose that there exists some $\bd Z$ such that $\bd Z \in \hat{\Omega}_{\tau}$ and $\bd Z \not\in \hat{\Omega}_{0}$. First, $\bd Z \in \hat{\Omega}_{\tau}$ implies that $ \forall l \in \{1, \ldots, L\}$ and $\forall m \in \{1, \ldots, M\}$ such that,
\begin{equation}
\label{eq:expre_prop2_dev1}
\bd 1_M^T\bd P_l\bd Z\bd c_m = 1,
\end{equation}
and 
\begin{equation}
\label{eq:expre_prop2_dev2}
\phi(\bd P_l\bd Z\bd c_m,\bd P_l \bm{\Pi} \bd c_m) \leq 1.
\end{equation}
Since $\bd Z \not\in \hat{\Omega}_{0}$, it implies that $\exists l \in \{1, \ldots, L\}$ and $\exists m \in \{1, \ldots, M\}$, such that $\bd P_l\bd Z\bd c_m \notin \{0,1\}^M$, i.e. it does not have a single entry equal to one and others being zero.

Set $\bd b = \bd P_l\bd Z\bd c_m$ and $\bm{\tau} = \bd P_l\bm{\Pi}\bd c_m$, we consider the following three cases:\\
(1) When there are more than one entries of $\bd b$ greater than $\bm{\tau}$, then $\phi(\bd b,\bm{\tau}) > 1$, violating Eq. (\ref{eq:expre_prop2_dev2}).\\
(2) When there are multiple entries of $\bd b$, all smaller than $\bm{\tau}$, according to the definition of the approximation function, we have that,
\begin{equation}
\phi(\bd b,\bm{\tau}) = \sum_{i=1}^M b_i/\tau_i.
\end{equation}
According to Eqs. (\ref{eq:expre_prop2_dev1}) and (\ref{eq:expre_prop2_dev2}), we have that
\begin{equation}
\phi(\bd b,\bm{\tau}) \leq \bd 1_M^T\bd b/\tau_{\text{min}} = 1/\tau_{\text{min}} \leq 1 \Rightarrow \tau_{\text{min}} \geq 1.
\end{equation}
where $\tau_{\text{min}}=\min{\tau_i,i=1,\ldots,M}$. Contradiction to $\bm{\tau} < 1$.\\
(3) When there is only one entry of $\bd b$ greater than the corresponding entry of $\bm{\tau}$, we assume $b_k > \tau_k, k \in \{1, \ldots, M\}$. Utilizing Eq. (\ref{eq:expre_prop2_dev1}), we have that
\begin{equation}
\sum_{i=1,i \neq k}^M b_i = 1-b_k.
\end{equation}
According to the definition of the approximation function and Eq. (\ref{eq:expre_prop2_dev2}), we further have that
\begin{equation}
\phi(\bd b,\bm{\tau}) = 1+(1-b_k)/\tau_k \leq 1 \Rightarrow b_k\geq 1.
\end{equation}
Combining with Eq. (\ref{eq:expre_prop2_dev1}), we further obtain that $b_k=1$ and $b_i=0, \forall i \neq k$. Contradiction to the assumption of $\bd P_l\bd Z\bd c_m \notin \{0,1\}^M$, or equivalently the assumption of $\bd Z \not\in \hat{\Omega}_{0}$. Thereby, $\bd Z \in \hat{\Omega}_{\tau} \Rightarrow \bd Z \in \hat{\Omega}_{0}$, i.e., $\hat{\Omega}_{\tau} \subseteq \hat{\Omega}_{0}$.


Assume that there exists some $\bd Z$ such that $\bd Z \in \hat{\Omega}_{b}$. That means there exists only one entry $p \in \{1+(l-1)M, \ldots, lM\}$ for each $l \in \{1, \ldots, L\}$ and $m \in \{1, \ldots, M\}$ such that $Z_{pm}=1$ and $Z_{im} = 0, \forall i \in \{1+(l-1)M, \ldots, lM\} \; \text{and} \; i\neq p$.  According to the property of the approximation function in Lemma 1, we have that
\begin{equation}
\phi(Z_{im}) = \begin{cases}
1 & i=p,\\
0 & i\neq p,
\end{cases}
\end{equation}
Thus, $\phi(\bd P_l \bd Z \bd c_m, \bd P_l \bm{\Pi} \bd c_m)=1,l=1,\ldots,L, m=1,\ldots,M$. That implies that $\bd Z \in \hat{\Omega}_{0} \rightarrow \bd Z \in \hat{\Omega}_{\tau}$, i.e., $\hat{\Omega}_{0} \subseteq \hat{\Omega}_{\tau}$. Thereby, we prove that $\hat{\Omega}_{0} = \hat{\Omega}_{\tau}$. In a nutshell, the problem $(AP_{\tau})$ is equivalent to the problem $(AP_{0})$.

\subsection{The proof of Eq. (\ref{eq:expre_subgradientZ})}
\label{subsec:proof_eq22}

Define $\bd b = \bd P_l\bd Z\bd c_m$ and $\bm{\tau} = \bd P_l\bm{\Pi}\bd c_m$. Based on Eq. (\ref{eq:expreb0}), $\phi(\bd b,\bm{\tau})$ is defined as,
\begin{align}
\label{eq:expre_phii}
\phi(\bd b,\bm{\tau})= \sum_{i=1}^M \phi(b_i,\tau) = \sum_{i=1}^M \frac{1}{\tau_i}[b_i - (b_i - \tau_i)^+], \nn
\end{align}
where $b_i = \bd r_i^T \bd P_l \bd Z \bd c_m$ and $\tau_i = \bd r_i^T \bd P_l \bm{\Pi} \bd c_m$. Proceeding from Eq. (\ref{eq:expre_phii}) and utilizing Lemma 1(3) and the chain rule, we have that,
\begin{align}
\frac{\partial \phi(\bd b, \bm{\tau})}{\partial \bd Z} & = \sum_{i=1}^M \frac{\partial \phi(b_i,\tau_i)}{\partial \bd Z},\\
&= \sum_{i=1}^M \frac{\partial \phi(b_i,\tau_i)}{\partial b_i} \cdot \frac{\partial b_i}{\partial \bd Z}, \nn\\
&= \sum_{i=1}^M g(b_i,\tau_i) (\bd P_l^T \bd r_i \bd c_m^T), \nn\\
&= \bd P_l^T \tilde{\bd r}_{m,l} \bd c_m^T, \nn
\end{align}
where $\tilde{\bd r}_{m,l} = \sum_{i=1}^M g(\bd r_i^T\bd P_l \bd Z \bd c_m,\bd r_i^T\bd P_l \bm{\Pi} \bd c_m) \bd r_i$ $\square$.

\subsection{Proof of Proposition 1}
\label{subsec:theorem2}

Assume that $\bd Z^{(k)} \in \hat{\Omega}_{\tau}$ is a feasible point for $(AP_{\tau})$, that is,
\begin{eqnarray}
&& \phi(\bd P_l \bd Z^{(k)} \bd c_m,\bd P_l \bm{\Pi} \bd c_m) \leq 1, l=1,\ldots,L, m=1,\ldots,M \nn\\
&& \bd 1^T_M \bd P_l \bd Z^{(k)} \bd c_m =1, l=1,\ldots,L, m=1,\ldots,M \nn\\ 
&& \bd Z^{(k)} \bd 1_M  = \bd 1_N. 
\end{eqnarray}
Proceeding from the third constraint of $(BP_{\tau})$, we can obtain that
\begin{equation}
\label{eq:expre_afsf}
\t{tr}\{(\bd c_m \tilde{\bd r}_{m,l}^T \bd P_l) (\bd Z - \bd Z^{(k)})\} \leq 0, l=1,\ldots,L, m=1,\ldots,M.
\end{equation}
From Eq. (\ref{eq:expre_subgradientZ}), we can see that $\tilde{\bd r}_{m,l}$ is relevant to the gradient of the approximation function $\phi(\bd P_l \bd Z^{(k)} \bd c_m,\bd P_l \bm{\Pi} \bd c_m)$. According to Lemma 1, the gradients corresponding to the ``one'' entries of $\bd Z^{(k)}$ are $0$, while the gradients corresponding to the ``zero'' entries of $\bd Z^{(k)}$ are $1/\tau$. In order to satisfy Eq. (\ref{eq:expre_afsf}), variable $\bd Z$ has to keep the ``zero'' entries of $\bd Z^{(k)}$. This implies that $\phi(\bd P_l \bd Z \bd c_m,\bd P_l \bm{\Pi} \bd c_m) \leq 1, l=1,\ldots,L, m=1,\ldots,M$ and the obtained solution $\bd Z^{(k+1)}$ is a feasible point for $(AP_{\tau})$. Therefore, once the initial search point $\bd Z^{(0)}$ is chosen feasible, all iterates are feasible and the problem $(BP_{\tau})$ will converge to the problem $(AP_{\tau})$.

\bibliographystyle{ieeetr}
\bibliography{mybiblio}

\end{document}